\documentclass[11pt]{article} 

\usepackage{fullpage}
\usepackage{multirow}
\usepackage{comment}
\usepackage{graphicx}
\usepackage{epstopdf}
\usepackage{amsmath}
\usepackage{cite}

\usepackage{amssymb}
\usepackage{setspace}
\usepackage{amsthm}
\usepackage{epsfig,color,colordvi,wasysym}
\usepackage[top=1in,bottom=1in,left=1in,right=1in]{geometry}

\author{Lei Meng\thanks{Department of Computer Science and Engineering, University of Notre Dame} \thanks{Interdisciplinary Center for Network Science and Applications
(iCeNSA)} \thanks{ECK Institute for Global Health} , 
Aaron Striegel\footnotemark[1] , and
Tijana Milenkovi\'{c}\footnotemark[1] \footnotemark[2] \footnotemark[3]
}

\begin{document}

\title{Local versus Global Biological Network Alignment}
\date{}

\maketitle

\begin{abstract}
Network alignment (NA) aims to find regions of similarities between
molecular networks of different species. There exist two NA
categories: local (LNA) or global (GNA). LNA finds small highly
conserved network regions and produces a many-to-many node mapping. GNA
finds large conserved regions and produces a one-to-one node
mapping. Given the different outputs of LNA and GNA, when a new NA
method is proposed, it is compared against existing methods from the
same category. However, both NA categories have the same goal: to
allow for transferring functional knowledge from well- to
poorly-studied species between conserved network regions. So, which
one to choose, LNA or GNA? To answer this, we introduce the first
systematic evaluation of the two NA categories. 

We introduce new measures of  
alignment quality that allow for fair comparison of the different LNA
and GNA outputs, as such measures do not exist. We provide
user-friendly software for efficient alignment evaluation that
implements the new and existing measures. We evaluate prominent LNA
and GNA methods on synthetic and real-world biological networks. We
study the effect on alignment quality of using different interaction
types and confidence levels.  We find that the superiority of one NA
category over the other is context-dependent. Further, when we
contrast LNA and GNA in the application of learning novel protein
functional knowledge, the two produce very different predictions,
indicating their complementarity. Our results and software provide
guidelines for future NA method development and evaluation.
\end{abstract}

\section{Introduction}
\subsection{Motivation, background, and related work}

With advancements of high throughput biotechnologies such as yeast
two-hybrid (Y2H) assays \cite{Yu2008} and affinity purification
coupled to mass spectrometry (AP/MS) \cite{Gstaiger2009}, large
amounts of protein-protein interaction (PPI) data have become
available \cite{KEGG2014,I2D2007,BIOGRID}. Comparative analysis
of PPI data across species is referred to as biological network
alignment (NA). NA is proving to be valuable, since it can be used to
transfer biological knowledge from well- to poorly-studied species,
thus leading to new discoveries in evolutionary biology.

\begin{figure}[h!]
\centering
(a)\includegraphics[width=0.2\linewidth]{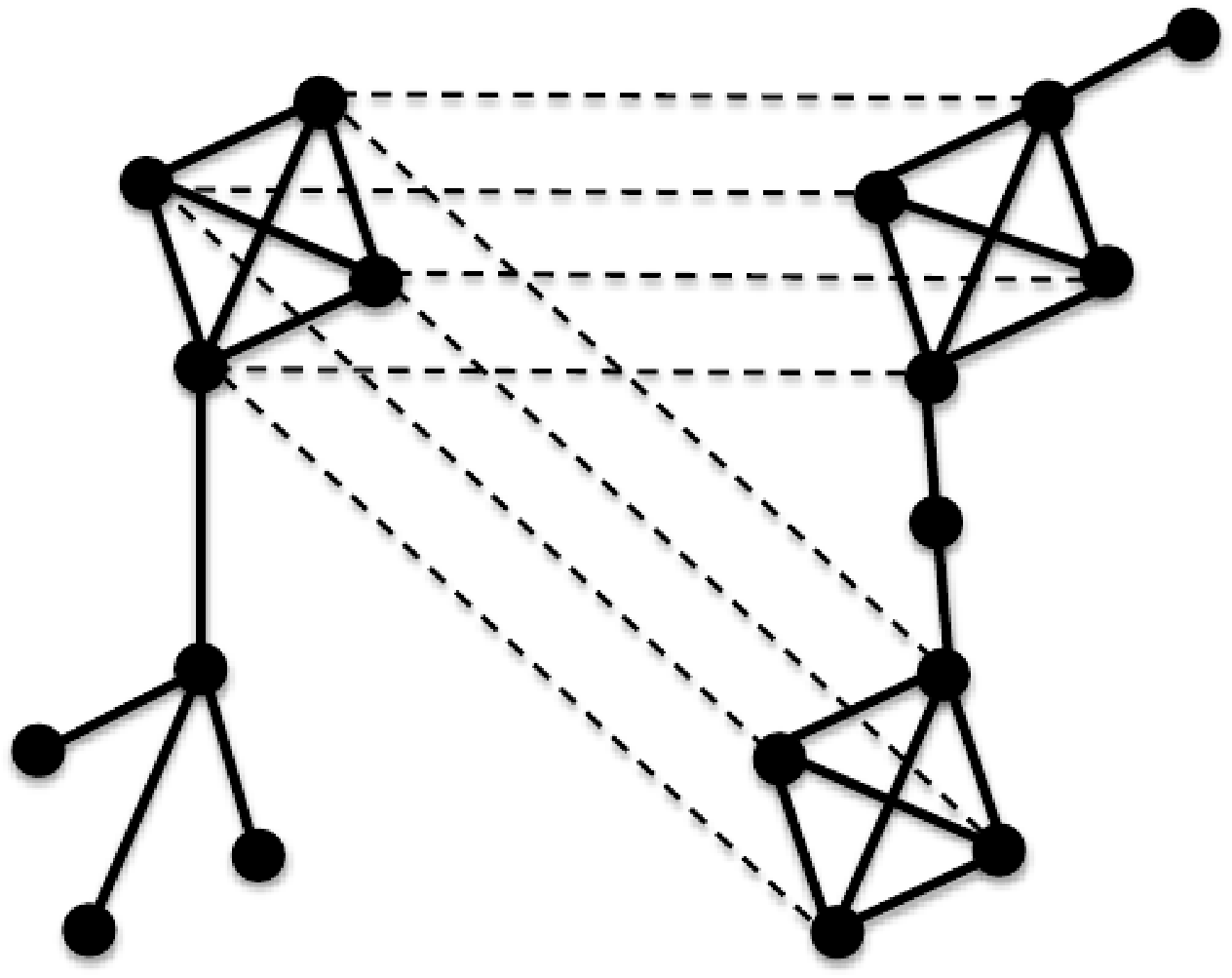}\hspace{1cm}
(b)\includegraphics[width=0.2\linewidth]{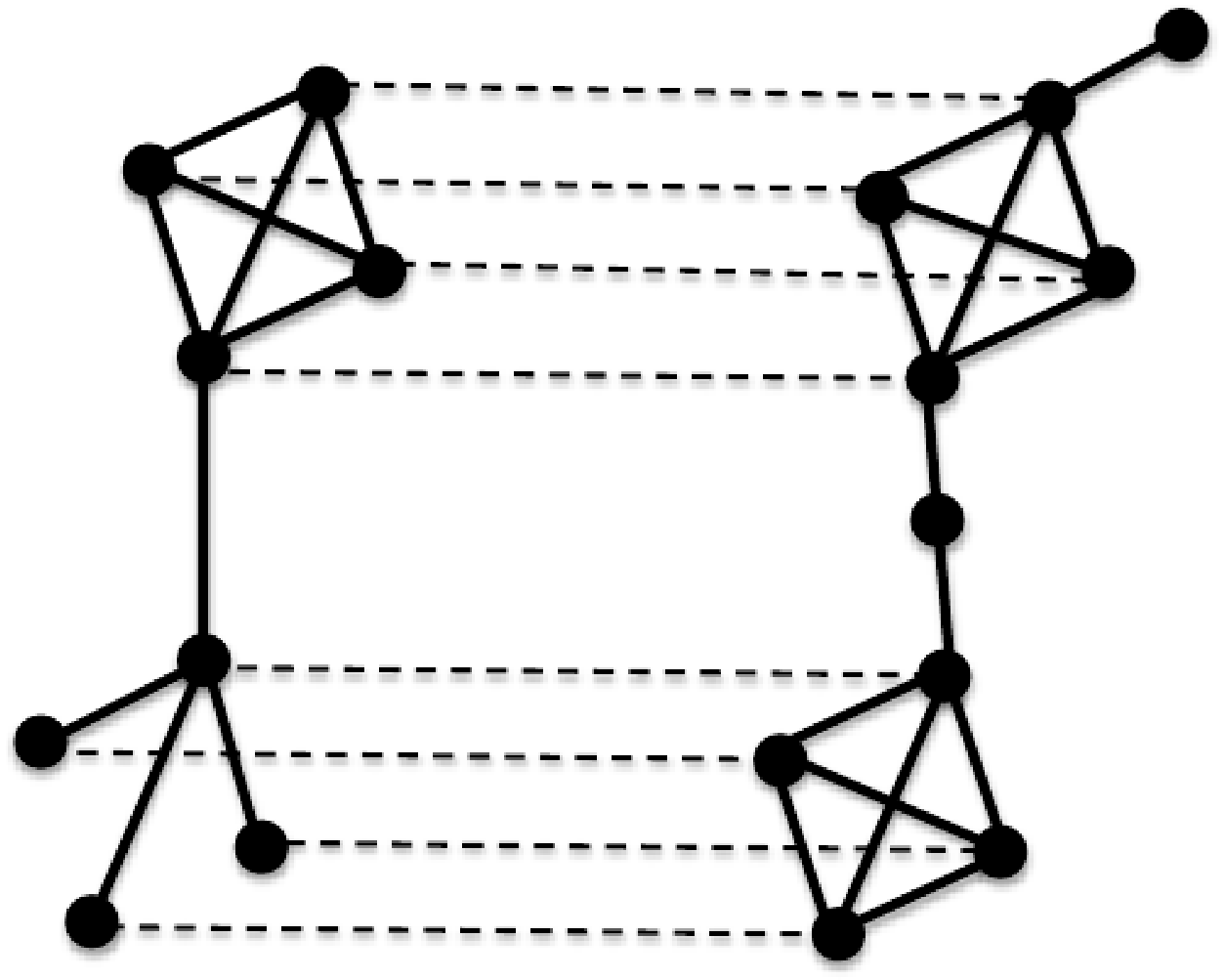}
\vspace{-0.2cm}
\caption{Illustration of \textbf{(a)} LNA and \textbf{(b)} GNA.  
}
\label{graph:NA}
\end{figure}

NA aims to find topologically and functionally similar (conserved)
regions between PPI networks of different species
\cite{Fazle2015}. Like genomic sequence alignment, NA can be
local (LNA) or global (GNA). LNA aims to find small highly conserved
subnetworks, irrespective of the overall similarity of compared
networks (Figure \ref{graph:NA} (a))
\cite{NETWORKBLAST,NETALIGNER,ALIGNNEMO,ALIGNMCL,LOCALALI}. Since
the highly conserved subnetworks can overlap, LNA results in
a many-to-many mapping between nodes of the compared networks -- a node
can be mapped to multiple nodes from the other network.  In contrast,
GNA aims to maximize overall similarity of the compared networks, at
the expense of suboptimal conservation in local regions (Figure
\ref{graph:NA} (b)). GNA  produces a one-to-one 
(injective) node mapping -- every node in the smaller network is
mapped to exactly one unique node in the larger network
\cite{IsoRank,HGRAAL,MIGRAAL,GHOST,NETAL,Todor2013,ibragimov2013gedevo,MAGNA,vijayan2015magna++,WAVE,LGRAAL,HUBALIGN,DualAligner}.

NA can also be categorized as \emph{pairwise} or \emph{multiple},
based on how many networks it can align.  See \cite{Fazle2015} for a
review of pairwise and multiple NA. Here, we focus on pairwise NA.

Given the different outputs of LNA and GNA, it is difficult to
directly compare them. Hence, when a new NA method is proposed, it is
compared only against existing methods from the same NA category. In
this context, NA methods can be evaluated with measures of topological
or biological alignment quality. An alignment is of good topological
quality if it reconstructs the underlying true node mapping well (when
this mapping is known) and if it conserves many edges. An alignment is
of good biological quality if the mapped nodes perform similar
function. LNA output is evaluated biologically but not
topologically. This is because LNA outputs a many-to-many node mapping 
and thus it is not clear how to compute edge conservation that has
been defined only for one-to-one mapping \cite{MAGNA}. GNA is
evaluated both topologically and biologically.

Despite the different output types of LNA and GNA, which makes their
direct comparison difficult, the two NA categories have the same ultimate
goal: to transfer functional knowledge from well- to poorly-studied
species, thus redefining the traditional notion of sequence-based
orthology to network-based orthology. For this reason, we introduce
the first ever  comparison of LNA and GNA.

\vspace{-0.2cm}

\subsection{Our approach and contributions}

\begin{figure*}[ht!]
\centering
\includegraphics[width=0.95\linewidth]{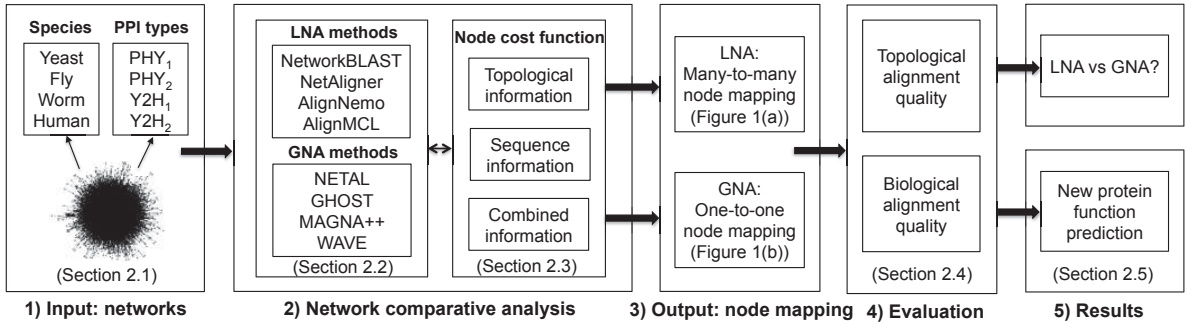}
\vspace{-0.4cm}
\caption{Summary of our LNA versus GNA evaluation framework, consisting 
of the following steps: \textbf{(1)} Input: networks from different
species containing different types of PPIs. Note that for this network
set, we do not know the true node mapping. Thus, we analyze an
additional set of networks with known true node mapping. \textbf{(2)}
Network comparative analysis: using prominent LNA or GNA methods (as
listed) to align networks across different species. During the
alignment construction process, we set each method's node cost
function (see Section \ref{sec:NCF}) to use topological information
only, sequence information only, or combined topological and sequence
information. \textbf{(3)} Output: many-to-many node mapping for LNA or
one-to-one node mapping for GNA. \textbf{(4)} Evaluation: measuring
topological and biological quality of each alignment. \textbf{(5)}
Results: fair comparison of LNA and GNA, and novel protein function
prediction.}
\label{graph:flowchart}
\end{figure*}

In the process of developing our novel framework for a fair comparison
of LNA and GNA (Figure \ref{graph:flowchart}), we do the following.

\noindent \textbf{1)} We evaluate eight prominent LNA and GNA methods.

\noindent \textbf{2)} We evaluate the NA methods on both synthetic 
networks with known true node mapping and real-world networks with
unknown true node mapping. For the latter, we explore the impact on
the results of using different PPI types
or PPIs of varying confidence.

\noindent \textbf{3)} We develop new alignment quality measures that 
allow for a fair comparison of LNA and GNA, since such measures do not
exist. We measure both topological and biological alignment quality.

\noindent \textbf{4)} We study the effect on the results of using only 
network topological information versus including also protein sequence
information into the alignment construction process.

\noindent \textbf{5)} Our LNA versus GNA evaluation reveals the following. 
When using only topological information during the alignment
construction process, GNA outperforms LNA both topologically and
biologically; when sequence information is also included, GNA is
superior to LNA in terms of topological alignment quality, while LNA
is superior to GNA in terms of biological quality. Different PPI types
and confidence levels lead to consistent findings, which indicates the
robustness of our approach to the choice of PPI data.

\noindent \textbf{6)} In addition to the thorough method evaluation, 
whose results provide guidelines for future NA method development,
we apply the NA methods to predict novel protein functional knowledge.
We find that LNA and GNA produce very different predictions,
indicating their complementarity when learning new biological
knowledge.

\noindent \textbf{7)} We provide a graphical user  interface (GUI) for 
NA evaluation integrating the new and existing alignment quality
measures.

\section{Methods}

\subsection{Data description}
\label{sec:datasets}

We analyze PPI networks with 1) known and 2) unknown true node
mapping.

\vspace{0.1cm}
\noindent\textbf{Networks with known true node mapping} 
contain a high-confidence \emph{S. cerevisiae} (yeast) PPI network
with 1004 proteins and 8323 PPIs \cite{Collins07} and five noisy
networks constructed by adding to the high-confidence network 5\%,
10\%, 15\%, 20\%, or 25\% of lower-confidence PPIs from the same
dataset \cite{Collins07}; the higher-scoring lower-confidence
PPIs are added first. We align the high-confidence network with each
of the noisy networks. Since all networks contain the same nodes, we
know the true node mapping. The high-confidence network is an exact
subgraph of each noisy yeast network. This popular evaluation test has
been adopted by many existing NA studies
\cite{GRAAL,HGRAAL,MIGRAAL,CGRAAL,GHOST,MAGNA,vijayan2015magna++}.

\vspace{0.1cm}
\noindent\textbf{Networks with unknown true node mapping} are 
PPI data from BioGRID (downloaded in November 2014) of four species:
\emph{S. cerevisiae} (yeast), \emph{D. melanogaster} (fly),
\emph{C. elegans} (worm), and \emph{H. sapiens} (human). For each
species, we extract four PPI networks containing different interaction
types and confidence levels: 1) all physical PPIs, where each PPI is
supported by at least one publication (PHY$_1$), 2) all physical PPIs,
where each PPI is supported by at least two publications (PHY$_2$), 3)
only yeast two-hybrid physical PPIs, where each PPI is supported by at
least one publication (Y2H$_1$), and 4) only yeast two-hybrid physical
PPIs, where each PPI is supported by at least two publications
(Y2H$_2$). We vary the PPI type (all physical interactions, most of
which are obtained by AP/MS, versus Y2H only) to test the robustness
of our approach to this parameter. We vary PPI confidence levels
because PPIs supported by multiple publications are more reliable than
those supported by only a single publication
\cite{cusick2009literature}.  For each network, we extract and
use its largest connected component (Supplementary Section S1 and
Supplementary Table S1).

\subsection{Network aligners evaluated in our study}\label{sec:aligners}

To evaluate LNA against GNA, we choose four prominent (pairwise) NA
methods from each of the LNA and GNA category.  More recent methods
are favored since they were shown to improve upon earlier
methods. Only methods with publicly available and relatively
user-friendly software are considered. As a result, we choose
NetworkBLAST \cite{NETWORKBLAST}, NetAligner
\cite{NETALIGNER}, AlignNemo \cite{ALIGNNEMO}, and AlignMCL
\cite{ALIGNMCL} from the LNA category, and we choose NETAL
\cite{NETAL}, GHOST \cite{GHOST}, MAGNA++
\cite{vijayan2015magna++}, and WAVE \cite{WAVE} from the GNA
category. An exception to the above guidelines is NetworkBLAST --
despite being an early LNA method, NetworkBLAST still remains a
popular LNA baseline. All methods are described in Supplementary
Section S2 and their parameters that we use are shown in Supplementary
Table S2.

\subsection{Aligners' node cost functions}
\label{sec:NCF}

All considered NA methods construct their alignments by first
computing pairwise similarities between nodes from different networks
via a node cost function (NCF). One can compute node similarities
by accounting for: 1) topological information only (T) in order to
measure how well the (extended) network neighborhoods of two nodes
match, 2) sequence information only (S) in order to measure the extent
of sequence conservation between the nodes, or 3) combined topological
and sequence information (T\&S).
We study the effect on alignment quality of using only topological
information versus also using sequence information in NCF. 

We evaluate each aligner for each of the three above cases. The exceptions
are NetworkBLAST, NetAligner, and NETAL, for the following reasons. Regarding NetworkBLAST and
NetAligner, they only allow for using sequence information within
NCF. The two methods require \emph{E}-value scores as input and it is
unclear how to convert topological information into values that are at
the same scale as the
\emph{E}-values. Regarding NETAL, its implementation failed to run
when we tried to include sequence information into its NCF. Topology-
and sequence-based NCFs that we use within the different NA methods
are discussed in Supplementary Section S3 and Supplementary Table
S3. Given topology- and sequence-based NCFs for two nodes from
different networks, we compute the nodes' combined (T\&S) NCF as the
linear combination of the individual NCFs: $NCF(T\&S)=\alpha\times
NCF(T)+(1-\alpha)\times NCF(S)$. We choose $\alpha=0.5$ to equally
balance between T and S.

\subsection{Evaluation of alignment quality}

\begin{figure}[h!]
\centering
(a)\includegraphics[width=0.1\linewidth]{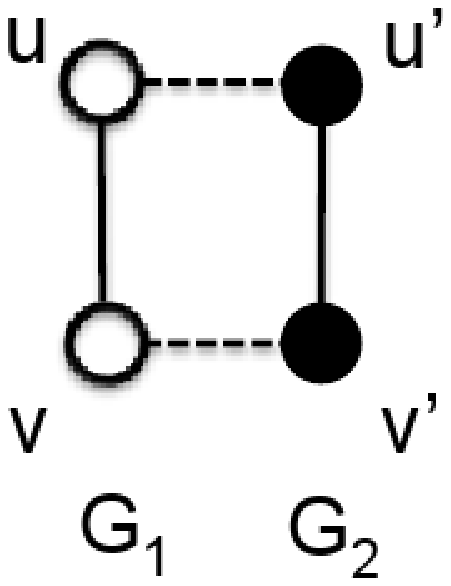}\hspace{1cm}
(b)\includegraphics[width=0.1\linewidth]{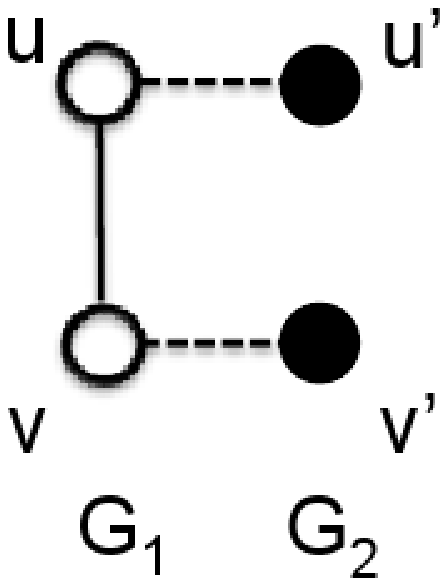}\hspace{1cm}
(c)\includegraphics[width=0.1\linewidth]{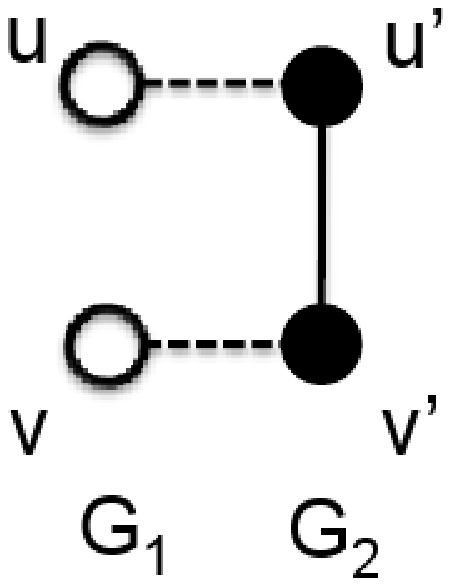}
\vspace{-0.3cm}
\caption{Illustration of \emph{conserved} and \emph{non-conserved} edges. 
\textbf{(a)} A conserved edge is formed by two edges $(u, v)\in G_1$ and 
$(u', v')\in G_2$ such that $u$ is aligned to $u'$ and $v$ is aligned
to $v'$.  A non-conserved edge is formed by \textbf{(b)} an edge $(u,
v)\in G_1$ and a non-edge $(u', v')\in G_2$ , or by \textbf{(c)} a
non-edge $(u,v)\in G_1$ and an edge $(u,v)\in G_2$, such that $u$ is
aligned to $u'$ and $v$ is aligned to $v'$. Nodes of the same color
come from the same network. A solid line represents an edge. Nodes
linked by a dashed line are aligned under $f$.}
\label{graph:quadruples}
\end{figure}

Next, we discuss measures that we use to evaluate topological (Section
\ref{sec:top_methods}) and biological (Section \ref{sec:bio_methods})
alignment quality. We introduce the following definitions. Let $f$ be
an alignment between two graphs $G_1(V_1, E_1)$ and $G_2(V_2,
E_2)$. Given a node $u$ from one graph, let $f(u)$ be the set of nodes
from the other graph that are aligned under $f$ to $u$. Given a node
set $V$, let $f(V)=\underset{v\in V}{\bigcup}f(v)$. Let $G'_1(V'_1,
E'_1)$ and $G'_2(V'_2, E'_2)$ be subgraphs of $G_1$ and $G_2$ that are
induced on node sets $f(V_2)$ and $f(V_1)$, respectively. We define
\emph{conserved} and \emph{non-conserved} edges as follows. A
conserved edge is formed by two edges from different networks such
that each end node of one edge is aligned under $f$ to a unique end
node of the other edge. In other words, a conserved edge is composed
of two edges from different networks that are aligned under $f$
(Figure \ref{graph:quadruples} (a)). A non-conserved edge is formed by
an edge from one network and a pair of nodes from the other network
that do not form an edge (i.e., that form a non-edge) such that each
end node of the edge is aligned under $f$ to a unique node of the
non-edge (Figure \ref{graph:quadruples} (b)-(c)).

\subsubsection{Topological evaluation}
\label{sec:top_methods}

First, we describe existing topological alignment quality measures,
along with their drawbacks. Next, we propose \emph{new} measures that
are motivated by the drawbacks of the existing measures.

\vspace{0.1cm}
\noindent\textbf{Existing measures.} Recall that intuitively an alignment 
is of high topological quality if it reconstructs the underlying true
node mapping well (when such mapping is known) and if it conserves
many edges. 

To evaluate how well an alignment reconstructs the true node mapping,
node correctness ($NC$) has been widely used
\cite{GRAAL,HGRAAL,MIGRAAL}. To date, $NC$ has
been defined only for GNA, as the fraction of nodes from the smaller
network that are correctly aligned (under \emph{injective} mapping
$f$) to nodes from the larger network with respect to the true node
mapping. The reason that $NC$ has not been defined for LNA is that
with LNA, a node from the smaller network can be mapped to multiple
nodes from the larger network, and thus, it is not clear how to
measure the percentage of nodes from the smaller network that are
correctly aligned. Hence, below, we generalize $NC$ for both LNA and
GNA.  $NC$ can only be used when the true node mapping is known.

To measure how well edges are conserved under an alignment, three
measures have been used to date: edge correctness ($EC$)
\cite{GRAAL}, induced conserved structure ($ICS$)
\cite{GHOST}, and symmetric substructure score ($S^3$)
\cite{MAGNA}. $S^3$ has been shown to be superior to $EC$ and
$ICS$, since intuitively it not only penalizes alignments from sparse
graph regions to dense graph regions (as $EC$ does), but also, it
penalizes alignments from dense graph regions to sparse graph regions
(as $ICS$ does). Hence, we only focus on $S^3$. Like $NC$, $S^3$ has
been only defined in the context of GNA, as
$\frac{|E^*_1|}{|E_1|+|E^{'}_2|-|E^*_1|}$, where $|E^*_1|$ is the
number of edges from $G_1$ that are conserved by $f$ (in this case,
$G_1$ is the smaller of the two networks in terms of the number of
nodes). The reason that $S^3$ has not been defined for LNA is that
with LNA that allows for many-to-many node mapping, it is not clear
how to count conserved edges, since an edge from one network could be
aligned to multiple edges from the other network. Hence, below, we
generalize $S^3$ to both LNA and GNA.

\vspace{0.1cm}
\noindent\textbf{New measures.}
To address the above issues, we propose new measures.

\noindent\textbf{1)} \emph{Precision, recall, and F-score of node correctness 
(P-NC, R-NC, and F-NC, respectively)}.  NC, defined only for GNA,
measures how well an alignment reconstructs the true node mapping. As
such, NC evaluates the 
\emph{precision} of the alignment --  the percentage of
the aligned node pairs that are also present in the true node
mapping. However, the corresponding \emph{recall} -- the percentage of
all node pairs from the true node mapping that are aligned under $f$
-- is not measured explicitly. This is because for GNA, recall has the
same value as precision. On the other hand, with LNA, precision and
recall could have different values. In order to generalize $NC$ for both
GNA and LNA, we propose  \emph{P-NC},
\emph{R-NC}, and \emph{F-NC}. Let $M$ be the set of node pairs that are mapped under
the true node mapping. Let $N$ be the set of node pairs that are
aligned under $f$. \emph{P-NC} is defined as $\frac{|M\cap
N|}{|M|}$. \emph{R-NC} is defined as $\frac{|M\cap
N|}{|N|}$. \emph{F-NC}, which combines \emph{P-NC} and \emph{R-NC}, is
the harmonic mean of the two individual measures. Like $NC$, our three
new measures can only be used when the true node mapping is known.

\noindent\textbf{2)} \emph{Generalized S$^3$ (GS$^3$)}. To generalize \emph{S$^3$} 
for both GNA and LNA, we propose \emph{GS$^3$} to count edge
conservation under $f$, independent on whether $f$ is injective or
many-to-many. We define \emph{GS$^3$} as the percentage of conserved
edges out of the total of both conserved and non-conserved edges. Let
$N_c$ and $N_n$ be the number of conserved and non-conserved edges,
respectively.
That is, $N_c=\sum_{uv\in E'_1}|\{(u', v')|u'\in f(u), v'\in f(v),
(u',v')\in E'_2\}|$.
$N_n$ is the sum of $N^1_{n}$ (i.e., the number of non-conserved edges
formed by aligning an edge from $G_1$ and a non-edge from $G_2$;
Figure \ref{graph:quadruples}b)) and $N^2_{n}$ (i.e., the number of
non-conserved edges formed by aligning a non-edge from $G_1$ and an
edge from $G_2$; Figure \ref{graph:quadruples}c)). In other words,
$N_n=N^1_{n}+N^2_{n}$, where $N^1_{n}$ and $N^2_{n}$ can be computed
as follows.
$N^1_{n}=N^{1'}_{n}-N_c$ and 
$N^2_{n}= N^{2'}_{n}-N_c$, where $N^{1'}_{n}=\sum_{uv\in E^{'}_{1}}|\{(u', v')| u'\in f(u), v'\in f(v),u'\neq v'\}|$ and $N^{2'}_{n}=\sum_{u'v'\in E^{'}_{2}}|\{(u, v)|u\in f(u'), v\in f(v'),u\neq v\}|$. Then, 
\vspace{-0.05cm}
$GS^3$ can be computed as: 
$GS^3=\frac{N_c}{N_c+N_n}=\frac{N_c}{N_c+N^1_n+N^2_n}=\frac{N_c}{N_c+(N^{1'}_n-N_c)+(N^{2'}_n-N_c)}=\frac{N_c}{N^{1'}_n+N^{2'}_n-N_c}$.
Clearly, for GNA, this formula for $GS^3$ is $S^3$ itself.

\noindent\textbf{3)} \emph{NCV combined with GS$^3$ (NCV-GS$^3$)}. 
Recall that \emph{GS$^3$} measures how well edges are conserved
between $G'_1$ and $G'_2$. LNA could produce small conserved
subgraphs, which could result in high $GS^3$ score. This would
mistakenly imply high alignment quality if we only rely on $GS^3$. But
if we adopt an additional criterion of what a good alignment is,
namely high node coverage (NCV), which is the percentage of nodes from
$G_1$ and $G_2$ that are also in $G'_1$ and $G'_2$ (i.e.,
$\frac{|V^{'}_1|+|V^{'}_2|}{|V_1|+|V_2|}$), then small conserved
subgraphs with high $GS^3$ would actually have low alignment quality
with respect to NCV. Thus, we combine \emph{NCV} and
\emph{GS$^3$} into \emph{NCV-G$S^3$} to get a more complete 
picture of the actual alignment quality. We define
\emph{NCV-G$S^3$}  as the geometric mean of
the two individual measures, because we want at least one low
alignment quality score to imply low combined score.

\subsubsection{Biological evaluation}
\label{sec:bio_methods}

To evaluate the biological quality of LNA and GNA, we use Gene
Ontology (GO) correctness \cite{GRAAL,MIGRAAL,NETAL} and the accuracy
of known protein function prediction
\cite{NETWORKBLAST,MIGRAAL,GHOST}.  Many GO annotations are
obtained via sequence comparison
\cite{Crawford2014FairEval}. Using such data to evaluate
alignments of NA methods that already use sequence information in NCF
would lead to biased results \cite{MIGRAAL}. Therefore, we only
use GO annotations that have been obtained experimentally.

\noindent\textbf{1)} \emph{GO correctness (GC)}. This measure quantifies 
the extent to which protein pairs that are aligned under $f$ are
annotated with the same GO terms (Supplementary Section S4)
\cite{GRAAL}.

\noindent\textbf{2)} \emph{Precision, recall, and F-score of known protein 
function prediction (P-PF, R-PF, and F-PF, respectively)}. We make GO
term prediction(s) for each protein from $G_1$ or $G_2$ that is
annotated with at least one GO term \cite{Fazle2014Aging} through a multi-step process. First,
we hide the protein's true GO term(s). Second, we predict its GO
term(s) based the GO term(s) of its aligned counterpart(s) under
$f$. After we make predictions for all proteins, we evaluate the 
precision, recall, and F-score of the prediction results with respect
to the true GO terms of the predicted proteins (Supplementary Section
S4).

\subsection{Application to novel protein function prediction}\label{sect:novel_predictions}

One application of NA is to predict \emph{novel} function of proteins
based on the annotations of their aligned counterparts under $f$. We
use LNA and GNA in this context to find statistically significant
alignments and make novel protein function predictions
from such alignments (Supplementary Section S5).

\section{Results and discussion}

We evaluate LNA against GNA on networks with known (Section
\ref{sect:results_known}) and  unknown (Section \ref{sect:results_unknown}) 
true node mapping.

\subsection{Networks with known true node mapping}\label{sect:results_known}

\subsubsection{Relationships between different alignment quality measures}
\label{yeast_correlation}
To fairly evaluate different NA methods, we first study relationships
and potential redundancies of different alignment quality measures in
order to select only non-redundant measures to fairly evaluate LNA
against GNA (Supplementary Section S6.1).

For networks with known true node mapping, we use the six topological
measures:
\emph{P-NC}, \emph{R-NC}, \emph{F-NC},
\emph{NCV}, \emph{GS$^3$}, and \emph{NCV-GS$^3$} (Section
\ref{sec:top_methods}).  That is, 
we do not use biological measures (which are \emph{approximate}
measures of similarity or correspondence between aligned nodes;
Section
\ref{sec:bio_methods}) because we  know the true node
mapping, i.e., the \emph{actual} correspondence between nodes that a
good aligner should be able to reconstruct well. For a given alignment
quality measure, we compute the score of aligning each of the five
pairs of networks with known true node mapping (Section
\ref{sec:datasets}) with each of the eight NA methods (Section
\ref{sec:aligners}). Then, for each pair of measures, we compute the Pearson
correlation coefficient across all of their $5 \times 8=40$ alignment
quality scores.  We do this for each type of information used within
NCF during alignment construction process, namely T, T\&S, and S
(Section
\ref{sec:NCF}).

\begin{figure}[h!]
\centering
(a)\includegraphics[width=0.25\linewidth]{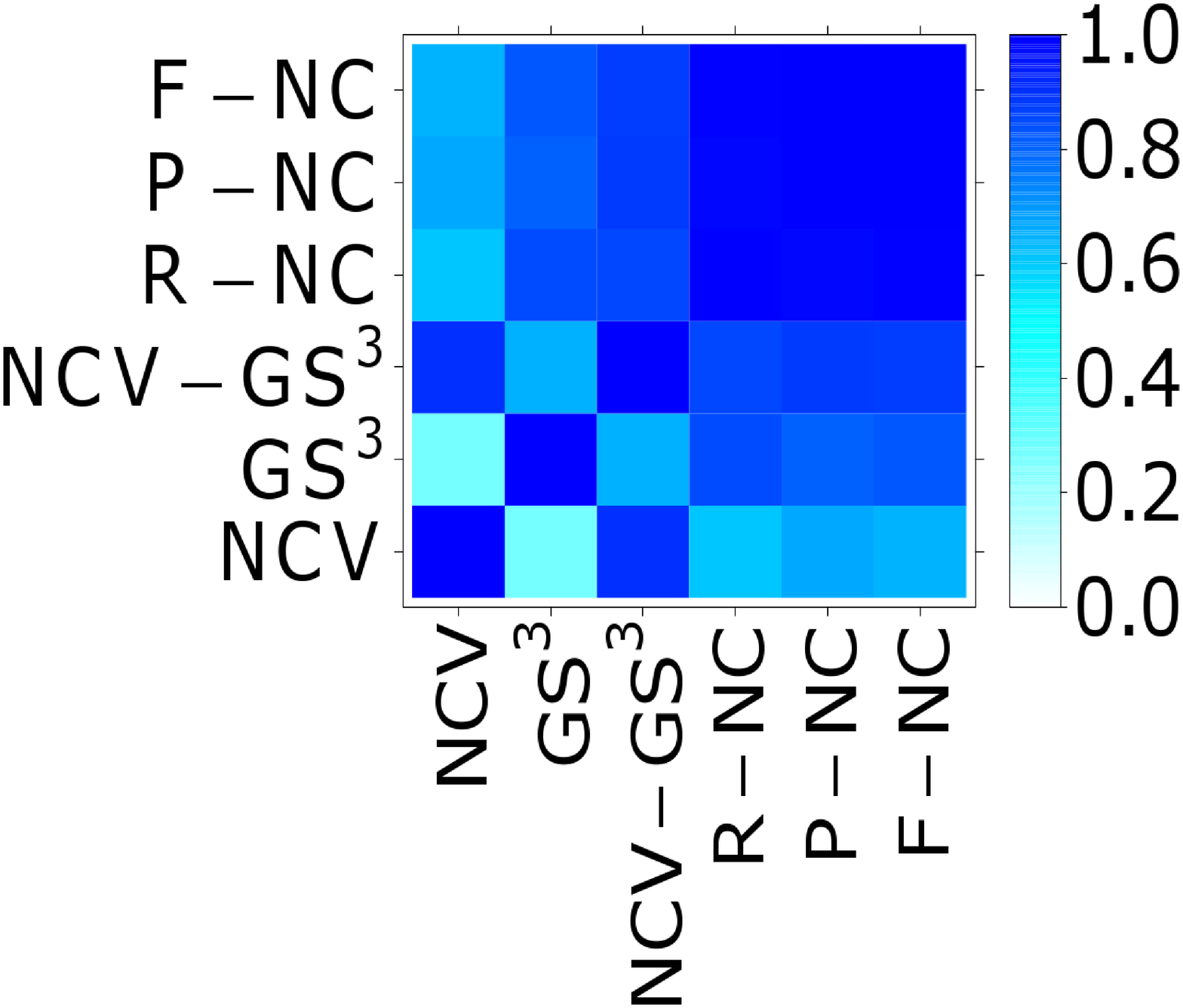}
(b)\includegraphics[width=0.25\linewidth]{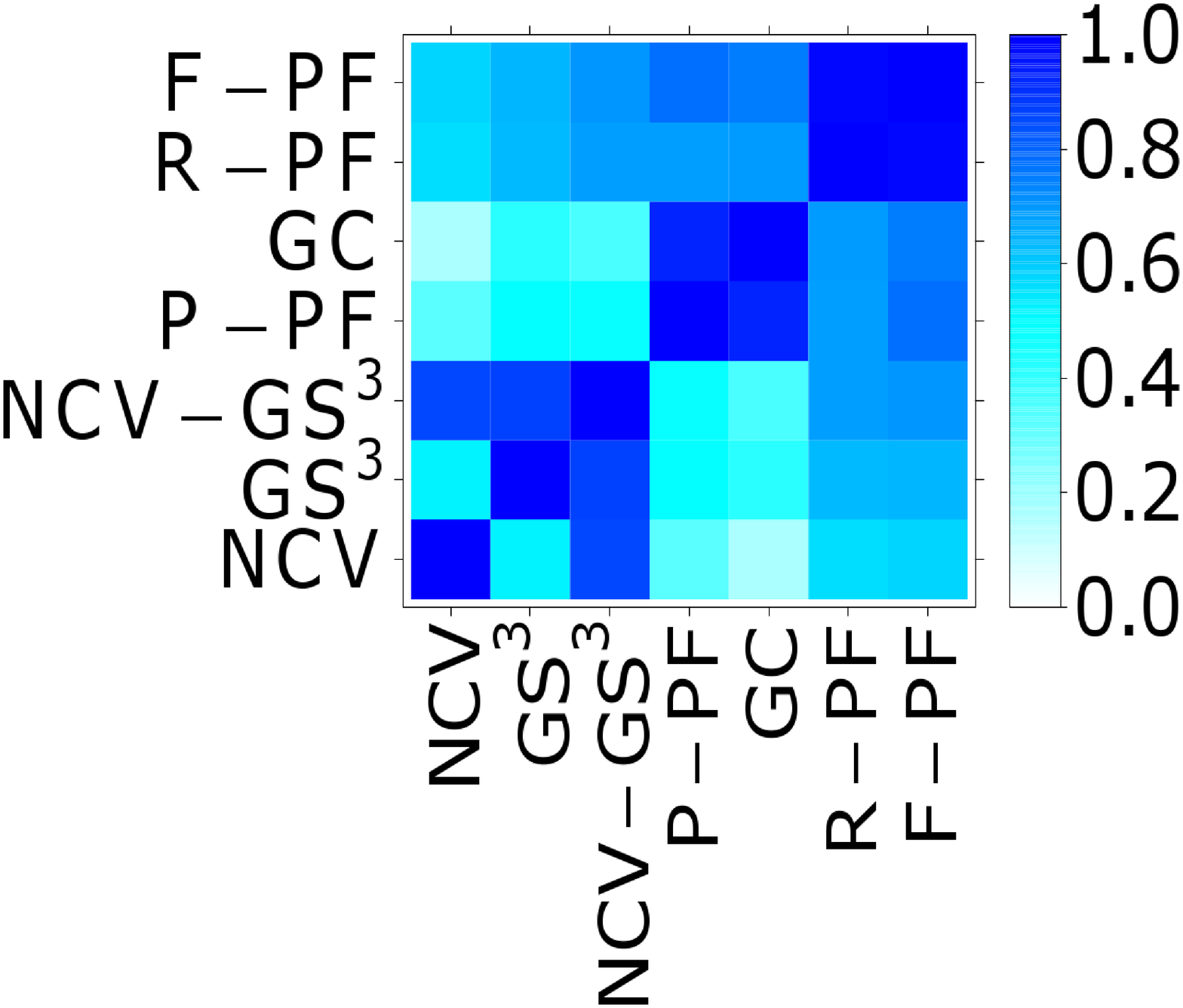}
\vspace{-0.3cm}
\caption{Pairwise relationships (Pearson correlations) between 
the six topological alignment quality measures over all alignments of
\textbf{(a)} networks with known true node mapping and \textbf{(b)}
networks with unknown true node mapping from four different species
(i.e., yeast, fly, worm and human) containing four different types of
PPIs (i.e., Y2H$_1$, Y2H$_2$, PHY$_1$, and PHY$_2$), for T. For T\&S
and S, see Supplementary Figures S1 and S5.}
\label{graph_measure_correlation}
\end{figure}

Since all six measures are topological, we expect them to be highly
correlated with each other. Indeed, this is what we observe: all pairs
of measures are significantly correlated independent of the type of
information used in NCF ($p$-values $< 10^{-3}$; Figure
\ref{graph_measure_correlation} (a) and Supplementary Figure S1). At the same time, since the six measures 
naturally cluster into two groups based on their definitions (one
group consisting of P-NC, R-NC, and F-NC, measures that quantify how
well the alignment captures the true node mapping, and the other group
consisting of NCV, GS$^3$, and NCV-GS$^3$, measures that capture the
size of the alignment), we expect within-group correlations to be
higher than across-group correlations.  Indeed, this is what we
observe (Figure
\ref{graph_measure_correlation} (a), Supplementary Section S6.1, and 
Supplementary Figure S1).
Since the two groups of measures evaluate alignment quality from
different perspectives, since within each group the measures are
redundant (Supplementary Section S6.1), and since in the first group
F-NC combines P-NC and R-NC while in the second 
group NCV-GS$^3$ combines NCV and GS$^3$,
henceforth, we focus on F-NC and NCV-GS$^3$ as the most
representative non-redundant measures.

\subsubsection{Comparison of LNA and GNA}
\label{yeast_comparison}

To fairly evaluate LNA against GNA, we perform ``all methods'' and
``best method'' comparisons of the NA methods. By ``all methods''
comparison, we mean the following: to claim that LNA is better than
GNA, each of the four LNA methods has to beat all four of the GNA
methods. Analogously, to claim that GNA is better than LNA, each of
the four GNA methods has to beat all four of the LNA methods. If none of the
two conditions are met, then we say that neither LNA nor GNA is
superior. By ``best method'' comparison, we mean the following: to
claim that LNA is better than GNA, at least one LNA method has to beat
all four of the GNA methods. Analogously, to claim that GNA is better than
LNA, at least one GNA method has to beat all four of the LNA methods. If none
of the two conditions are met, then we say that neither LNA nor GNA is
superior.  We perform each of the ``all methods'' and ``best method''
comparisons with respect to each of T, T\&S, S, and B, where B is the
best-case scenario, i.e., the best of T, T\&S, and S. Namely, given
two networks and a NA method, three alignments will be produced, one
for each of T, T\&S, and S. Then, B is the best of the three
alignments with respect to the given alignment quality measure
(different alignment quality measures might identify different
alignments as B out of T, T\&S, and S).

\begin{figure}[h!]
\centering
(a)\includegraphics[width=0.2\linewidth]{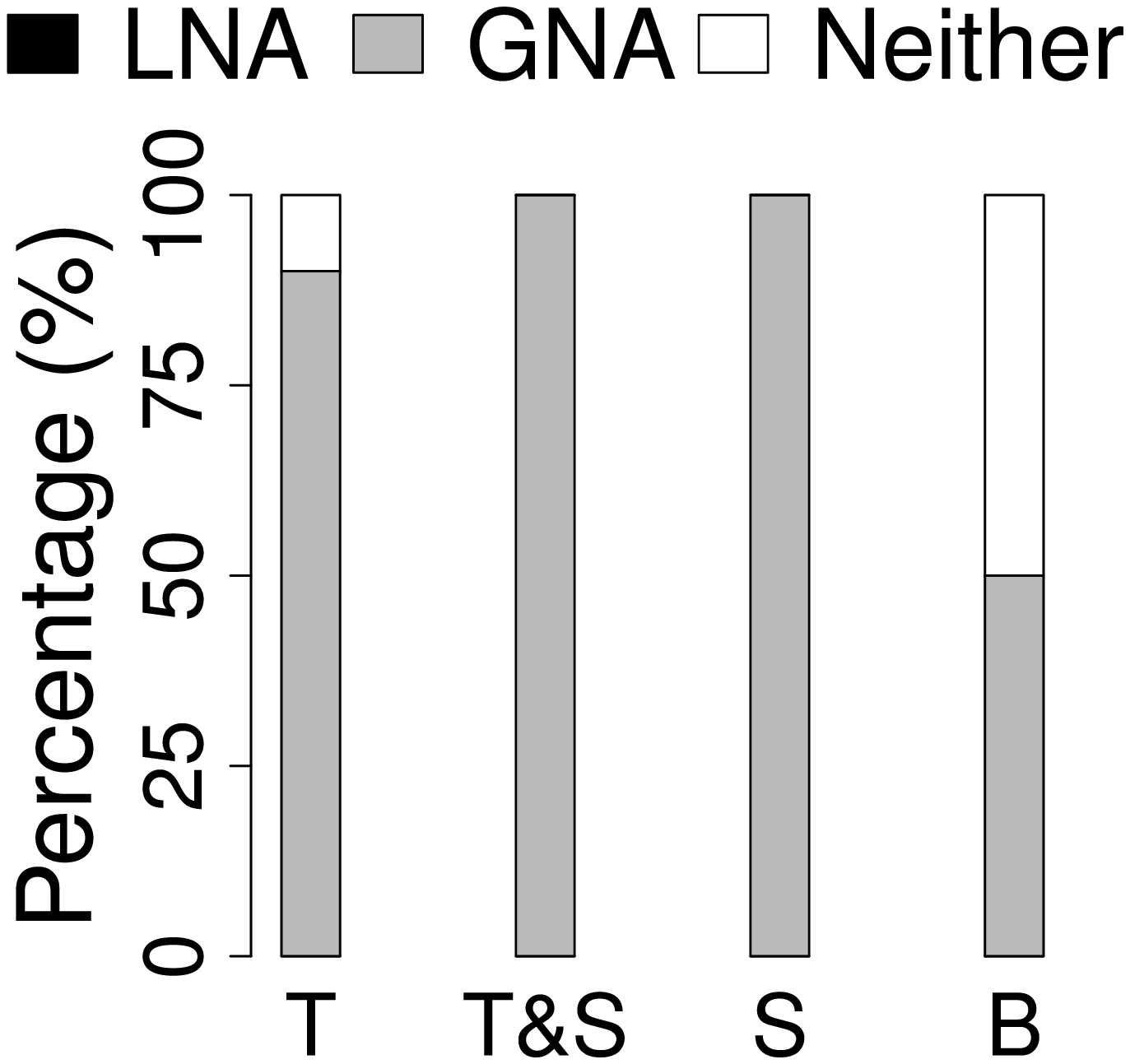}
\hspace{1cm}
(b)\includegraphics[width=0.2\linewidth]{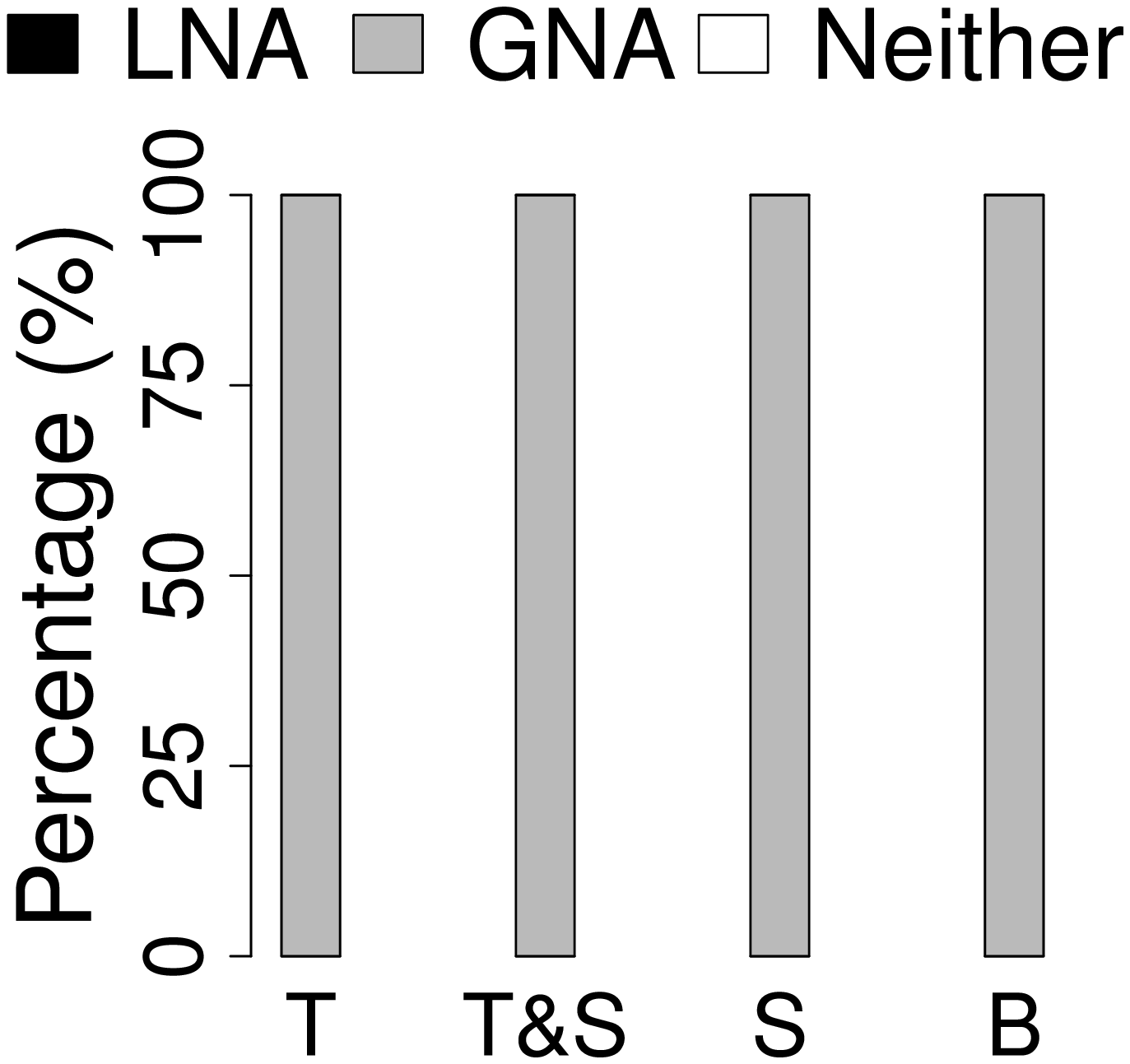}
\vspace{-0.35cm}
\caption{Overall comparison of LNA and GNA for networks with known true node 
mapping with respect to \textbf{(a)} \emph{``all methods''} and
\textbf{(b)} \emph{``best method''} comparison, for T,
T\&S, S, and B. Each bar shows the percentage of the aligned network
pairs (over both considered alignment quality measures combined) for
which LNA is superior (black), GNA is superior (grey), or neither LNA
nor GNA is superior (white). For detailed results, see Figure
\ref{graph:yeast_lineplot_T} and Supplementary Figures S3 and S4.}
\label{graph_yeast_group}
\end{figure}

Overall, for both ``all methods'' and ``best method'' comparisons, we
observe that GNA is superior to LNA in most of the cases for each of
T, T\&S, S, and B (Figure \ref{graph_yeast_group} and Supplementary Figure S2).

\begin{figure}[h!]
\centering
(a)\includegraphics[width = 0.2\linewidth]{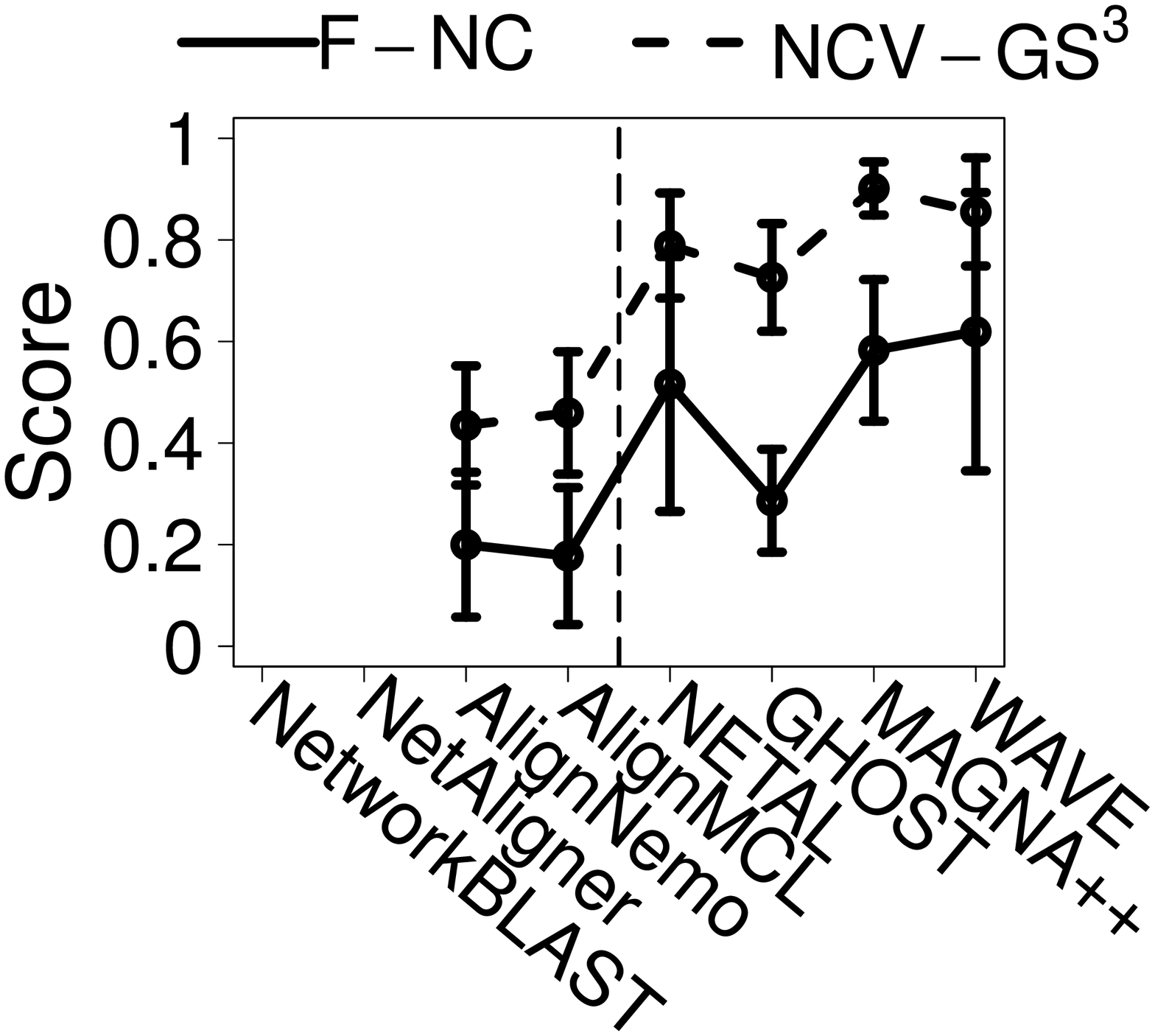}
(b)\includegraphics[width = 0.2\linewidth]{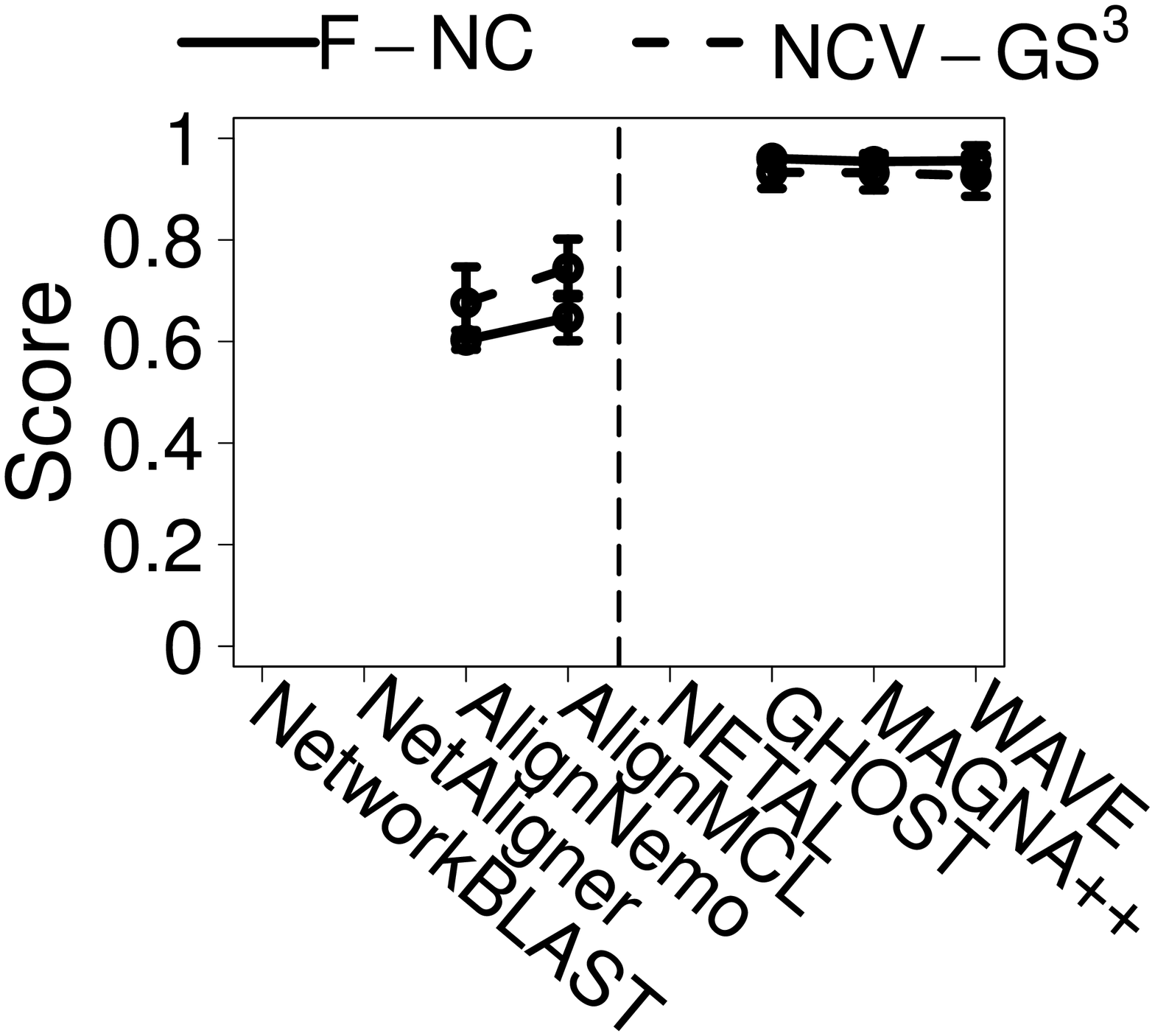}
(c)\includegraphics[width = 0.2\linewidth]{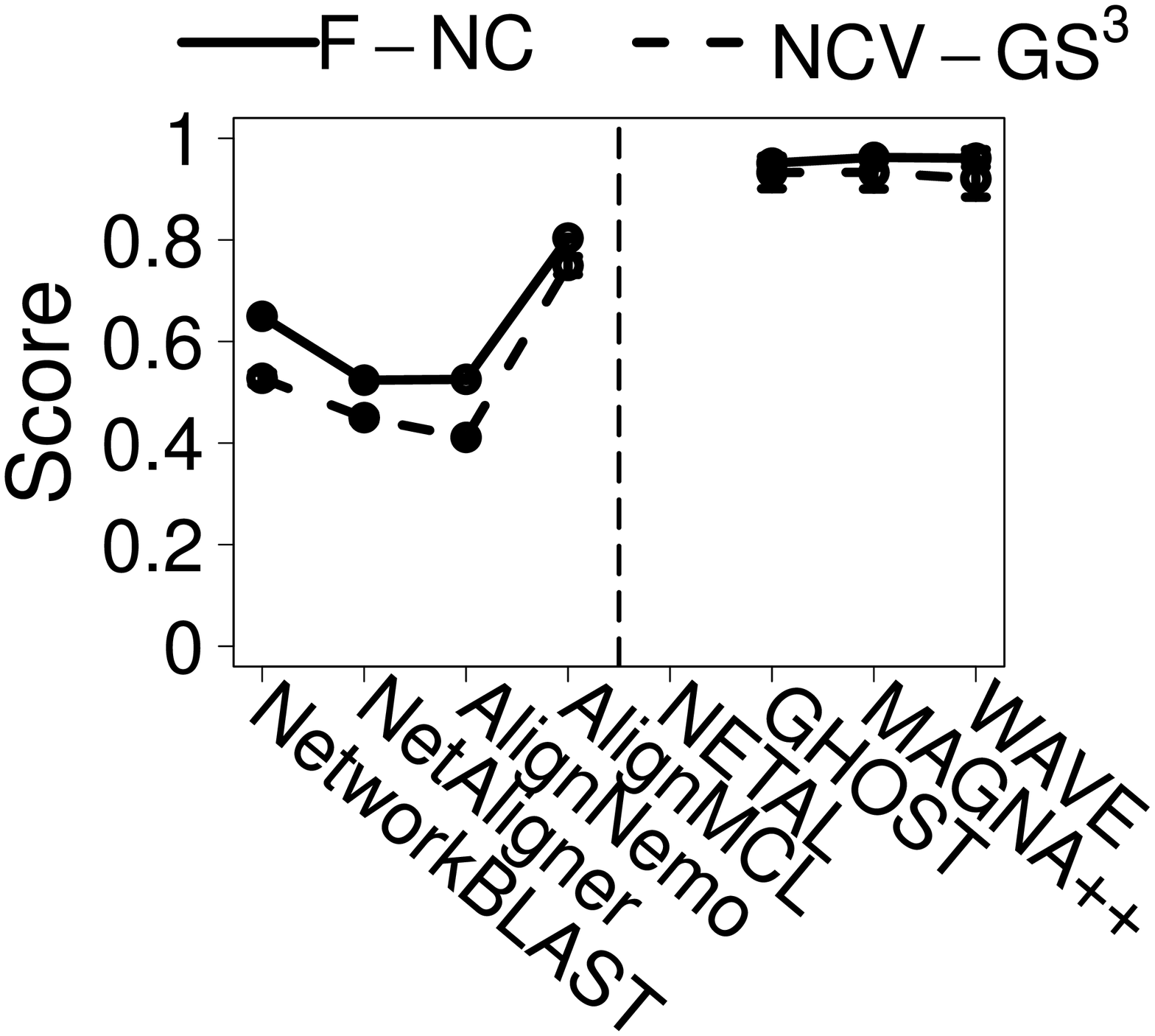}
(d)\includegraphics[width = 0.2\linewidth]{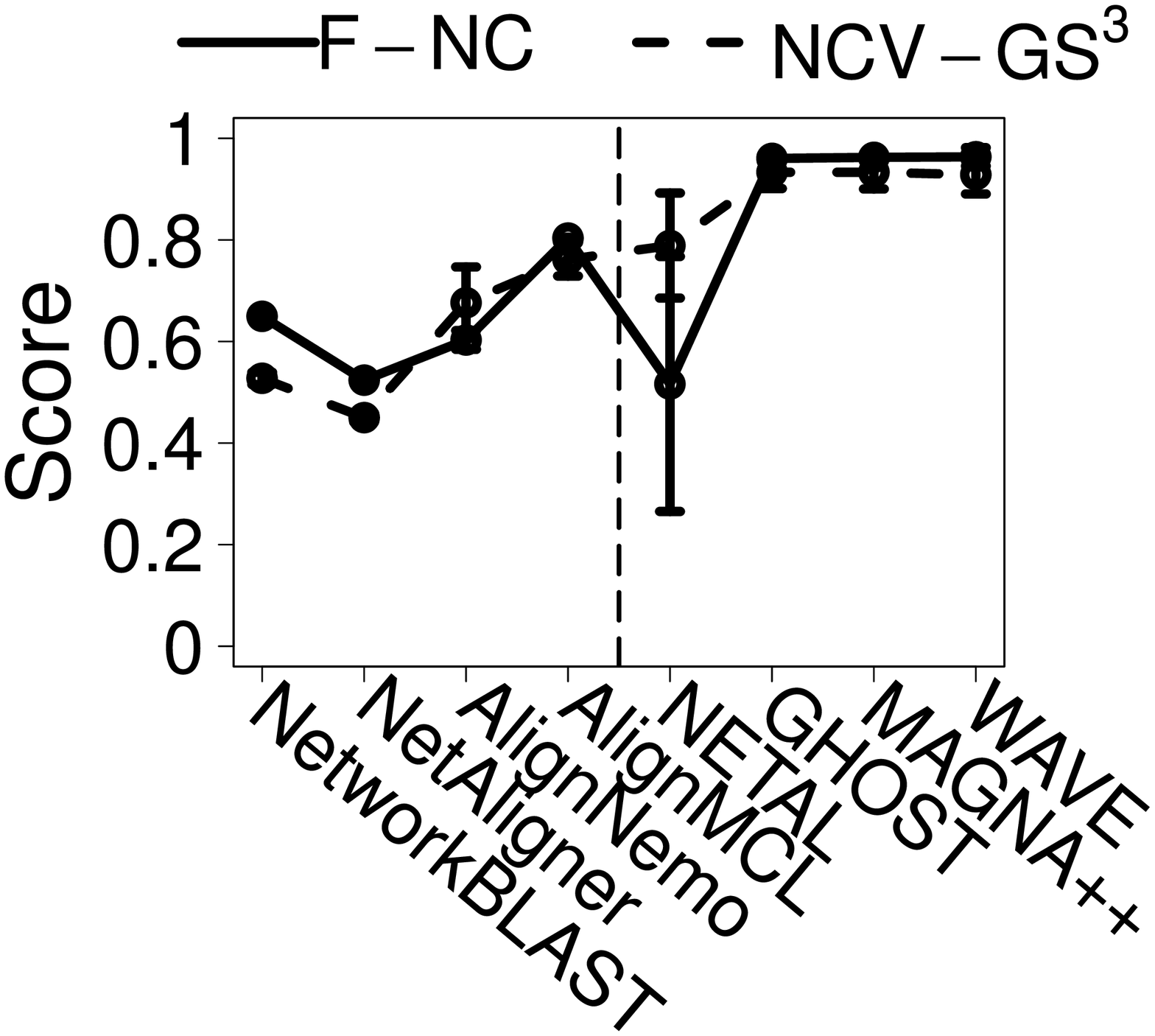}
\vspace{-0.3cm}
\caption{Detailed comparison of LNA and GNA for networks with known true 
node mapping with respect to F-NC and NCV-GS$^3$ alignment quality
measures, for \textbf{(a)} T, \textbf{(b)} T\&S,
\textbf{(c)} S, and \textbf{(d)} B. Each point
represents alignment quality of the given NA method averaged over all
network pairs, and each bar represents the corresponding standard
deviation. A missing point indicates that the given NA method cannot
use the corresponding type of information in NCF and thus no result is
produced.}
\label{graph:yeast_lineplot_T}
\end{figure}

Next, we zoom into the results (Figure
\ref{graph:yeast_lineplot_T} and Supplementary Figures S3 and S4)  to identify the best method(s). 
Recall that for LNA, NetworkBlast and NetAligner do not allow for
using topological information in NCF. Thus, we cannot consider these
methods for T and T\&S. Given this, the results for LNA are as
follows. For T and T\&S, the remaining methods, AlignNemo and
AlignMCL, are comparable (Figure
\ref{graph:yeast_lineplot_T} (a) and (b)). For S and  B, 
of all four methods, AlignMCL is superior (Figure
\ref{graph:yeast_lineplot_T} (c) and (d)). Hence, we conclude AlignMCL 
to be the best LNA method.  Recall that for GNA, NETAL does not allow
for using sequence information in NCF. So, we cannot consider this
method for T\&S and S. (For this method, B is the same as T).  Given
this, the results for GNA are as follows. For all of T, T\&S, S, and
B, WAVE and MAGNA++ are the best methods. GHOST also performs well
whenever sequence is used in NCF, i.e., for T\&S, S, and B. However,
for T, GHOST is inferior to all other methods. Hence, we conclude WAVE
and MAGNA++ to be the best GNA methods.  For both LNA and GNA, the
choice of aligned networks impacts the results less for T\&S and S
than for T (as supported by smaller standard deviations in Figure
\ref{graph:yeast_lineplot_T} (b) and (c) compared to  Figure
\ref{graph:yeast_lineplot_T} (a)).

\subsubsection{Summary}

For networks with known true node mapping, GNA is superior to LNA.
AlignMCL is superior of all LNA methods. WAVE and MAGNA++ are superior
of all GNA methods.

\subsection{Networks with unknown true node mapping}\label{sect:results_unknown}

\subsubsection{Relationships of different alignment quality measures}
\label{}

Similar to our analysis for networks with known true node mapping
(Section
\ref{yeast_correlation}), our first goal for four sets of networks
with unknown true node mapping (Y2H$_1$, Y2H$_2$, PHY$_1$, and
PHY$_2$, which encompass different species, PPI types, and PPI
confidence levels; Section
\ref{sec:datasets}) is to understand potential redundancies of
different alignment quality measures and choose the best and most
representative of all redundant measures for fair evaluation of LNA
and GNA.  All reported results are for all four sets of networks
combined, unless otherwise noted. In Section \ref{sect:robust}, we
break down the results per network set, in order to evaluate their
robustness to the choice of network data in terms of PPI type and
confidence level.

For the networks with unknown node mapping, we use all seven measures:
three topological (\emph{NCV}, \emph{GS$^3$}, and \emph{NCV-GS$^3$};
Section
\ref{sec:top_methods}) and  four biological (\emph{P-PF}, \emph{R-PF}, 
\emph{F-PF}, and \emph{GC}; Section \ref{sec:bio_methods}). 
For a given measure, we compute the score of aligning each of the 14
pairs of networks with known true node mapping (Section
\ref{sec:datasets}) with each of the eight NA methods (Section
\ref{sec:aligners}). Then, for each pair of measures, we compute the
Pearson correlation coefficient across all of their $14
\times 8=112$ alignment quality scores.  We do this for each type
of information used within NCF during alignment construction process,
namely T, T\&S, and S (Section \ref{sec:NCF}).

Since the seven measures naturally cluster into two groups (one group
consisting of the three topological measures that capture the size of
the alignment in terms of the number of nodes or edges, and the other
group consisting of the four biological measures that quantify the
extent of functional similarity of the aligned nodes), we expect
within-group correlations to be higher than across-group
correlations. Indeed, this is what we observe overall for all of T,
T\&S, and S (Figure \ref{graph_measure_correlation} (b) and
Supplementary Figure S5; also, see Supplementary Section S6.2 for more
details). Thus, since the two groups of measures evaluate alignment
quality from different perspectives, since within each group the
measures are redundant (Supplementary Section S6.2), and since in the
first group NCV-GS$^3$ combines
NCV and GS$^3$ (Section \ref{sec:top_methods}) while 
in the second group F-PF combines P-PF and R-PF
(and is also redundant to the remaining GC measure), henceforth, we
focus on NCV-GS$^3$ as the most representative non-redundant
topological measure and on F-PF as the most representative
non-redundant biological measure.

\subsubsection{Comparison of LNA and GNA} 
\begin{figure}[h!]
\centering
(a)\includegraphics[width=0.2\linewidth]{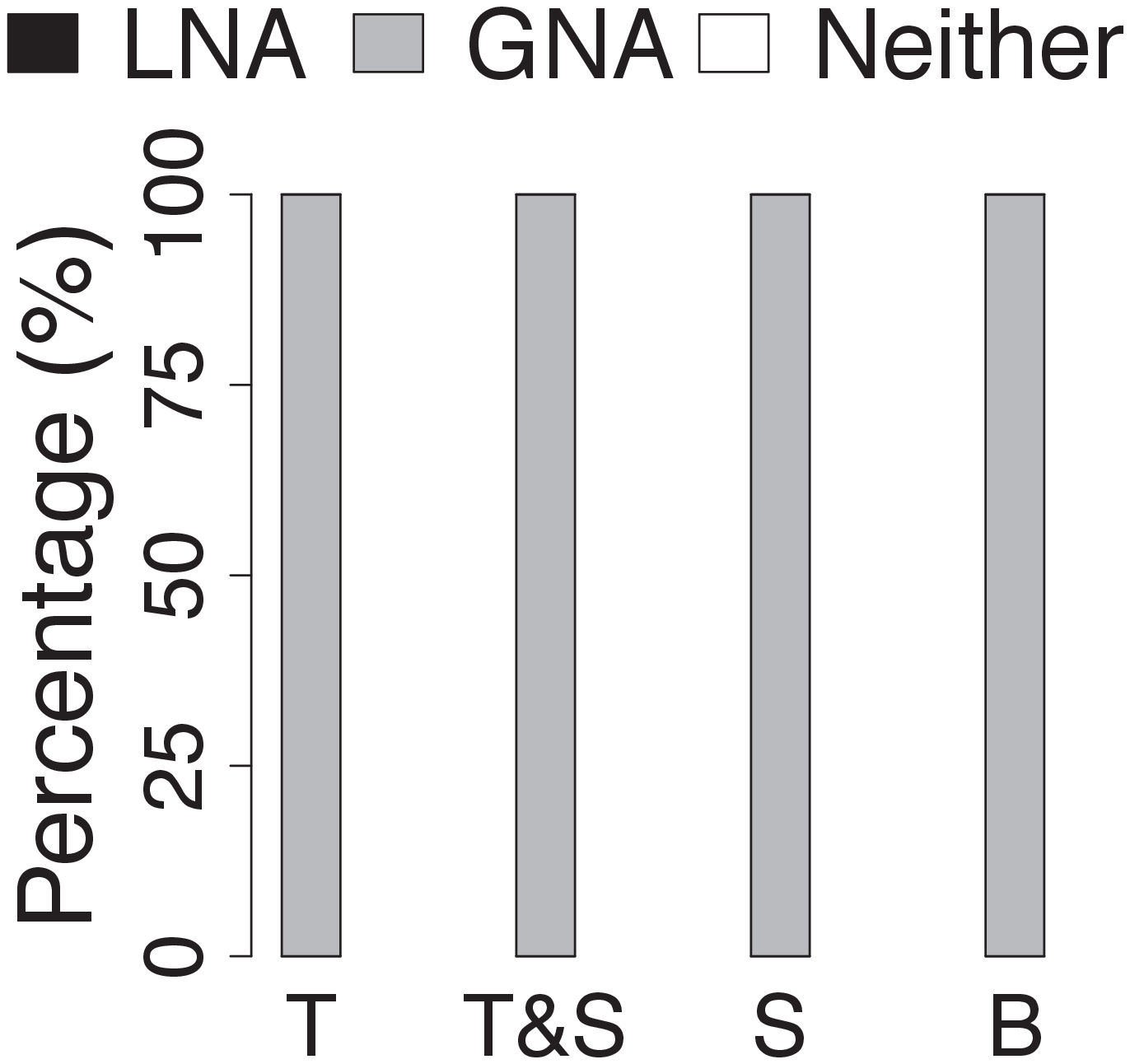}\hspace{0.5cm}
(b)\includegraphics[width=0.2\linewidth]{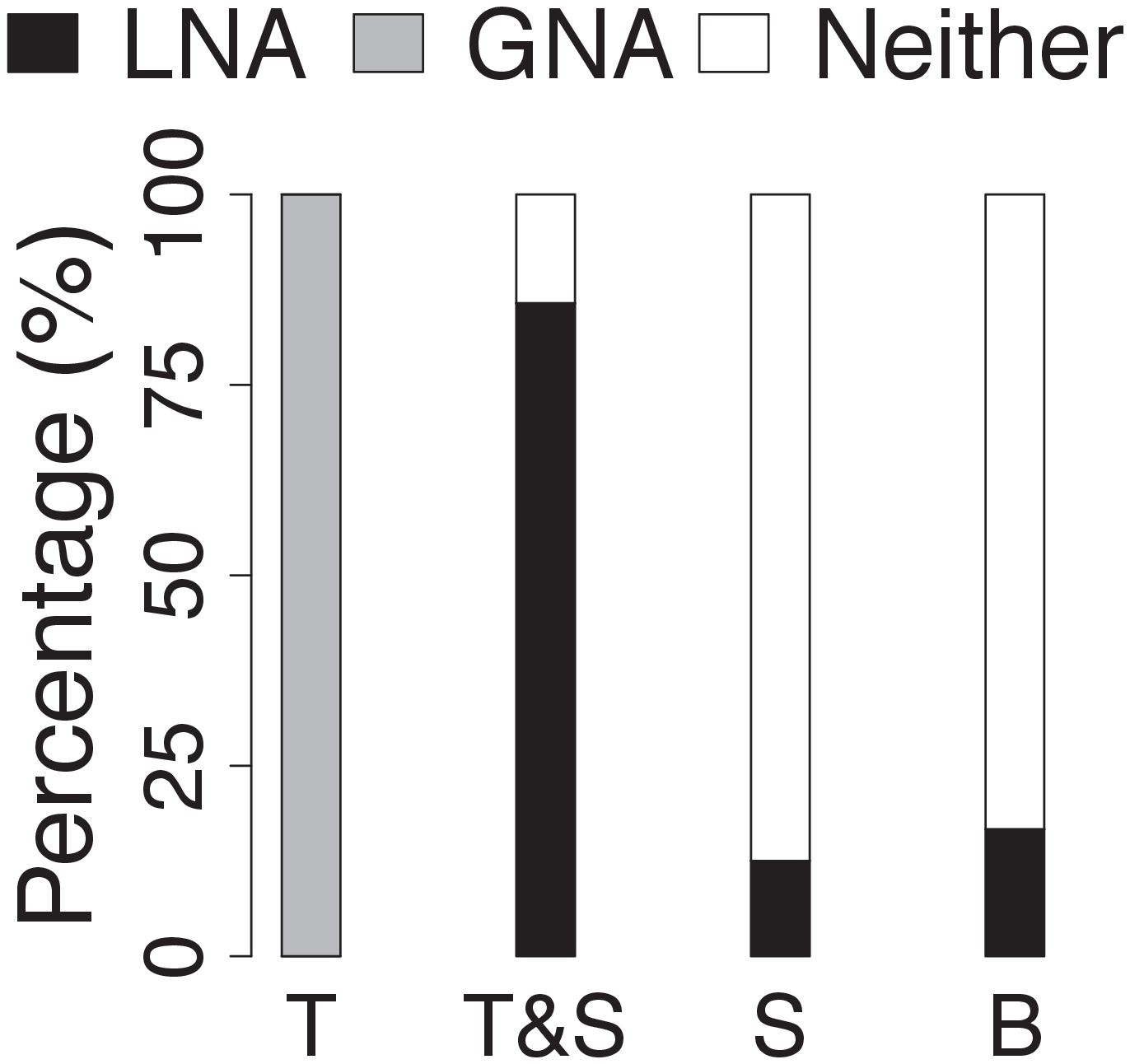}
\vspace{-0.3cm}
\caption{Overall comparison of LNA and GNA for networks with unknown 
true node mapping from four different
species (i.e., yeast, fly, worm and human) containing four different types of PPIs 
(i.e., Y2H$_1$, Y2H$_2$, PHY$_1$, and PHY$_2$) 
with respect to \textbf{(a)} \emph{``all methods''}
comparison and the topological NCV-GS$^3$ measure and \textbf{(b)}
\emph{``all methods''} comparison and the biological
 F-PF measure. Results are
shown for T, T\&S, S, and B. Each bar shows the percentage of the
aligned network pairs for which LNA is superior (black), GNA is
superior (grey), or neither LNA or GNA is superior (white). For ``best methods'' comparison results, see Supplementary Figure S6.}
\label{graph_biogrid_overall_comparison}
\end{figure}

As in our analysis from Section \ref{yeast_comparison}, here we
perform ``all methods'' and ``best method'' comparisons. For both
comparison types, with respect to topological alignment quality, GNA
is always superior to LNA for each of T, T\&S, S, and B (Figure
\ref{graph_biogrid_overall_comparison} (a), Supplementary Figure S6
(a), and Supplementary Figure S7). With respect to biological alignment quality, GNA is superior to
LNA for T, while LNA is comparable or superior to GNA for T\&S, S, and
B (Figure
\ref{graph_biogrid_overall_comparison} (b), Supplementary Figure S6
(b), and Supplementary Figure S7).

\begin{figure*}[Ht!]
\centering
(a)\includegraphics[width=0.21\linewidth]{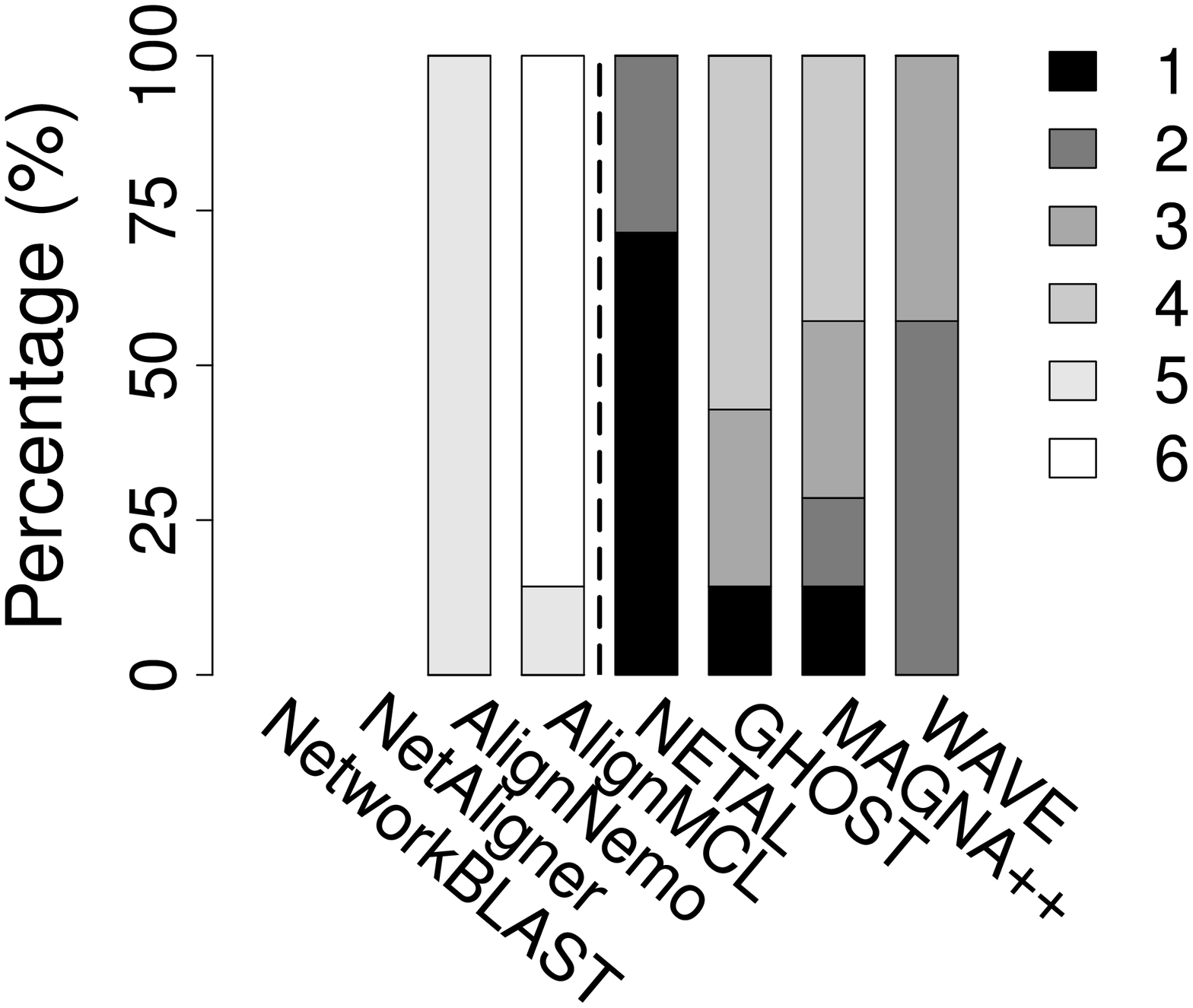}
(b)\includegraphics[width=0.21\linewidth]{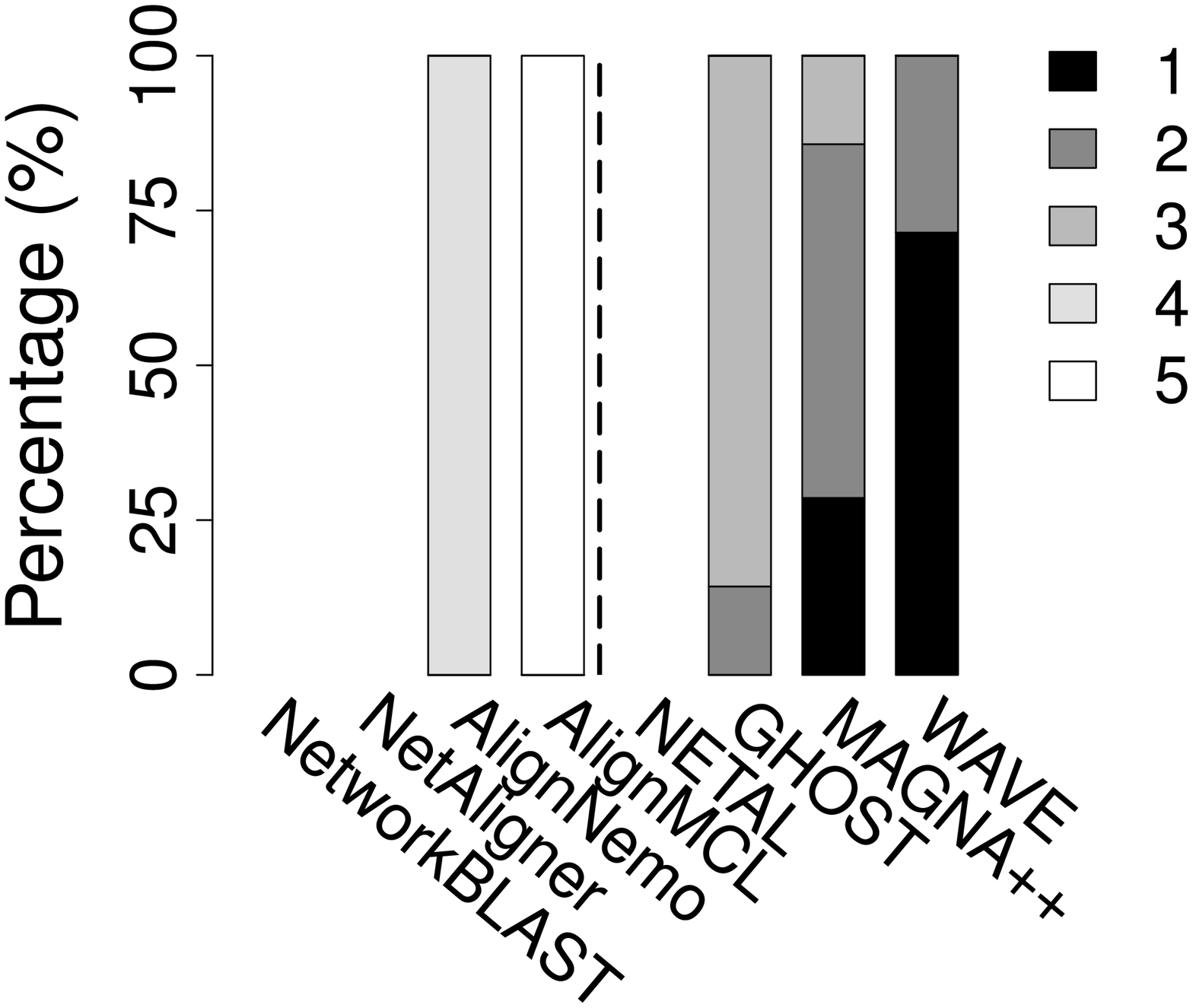}
(c)\includegraphics[width=0.21\linewidth]{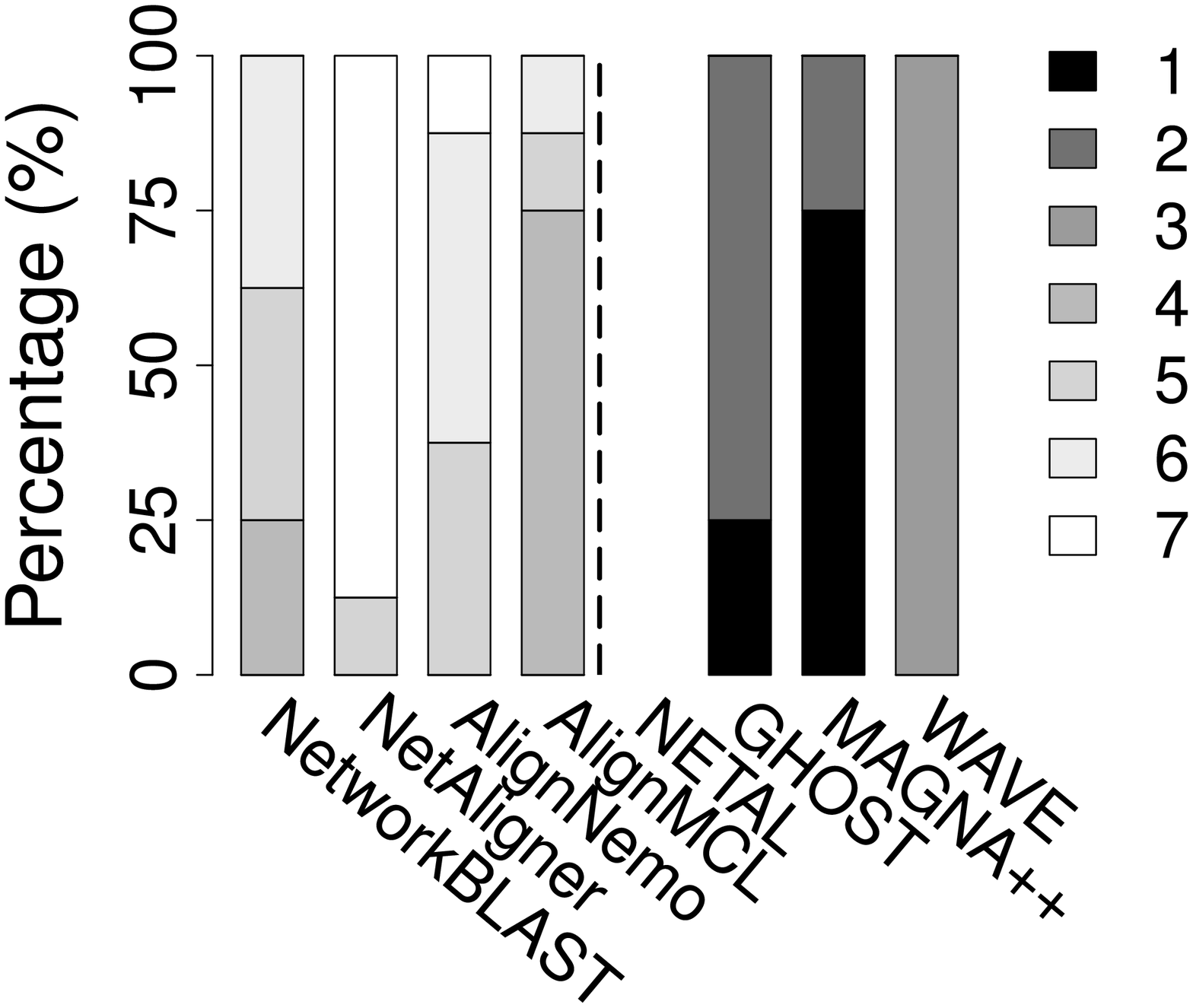}
(d)\includegraphics[width=0.21\linewidth]{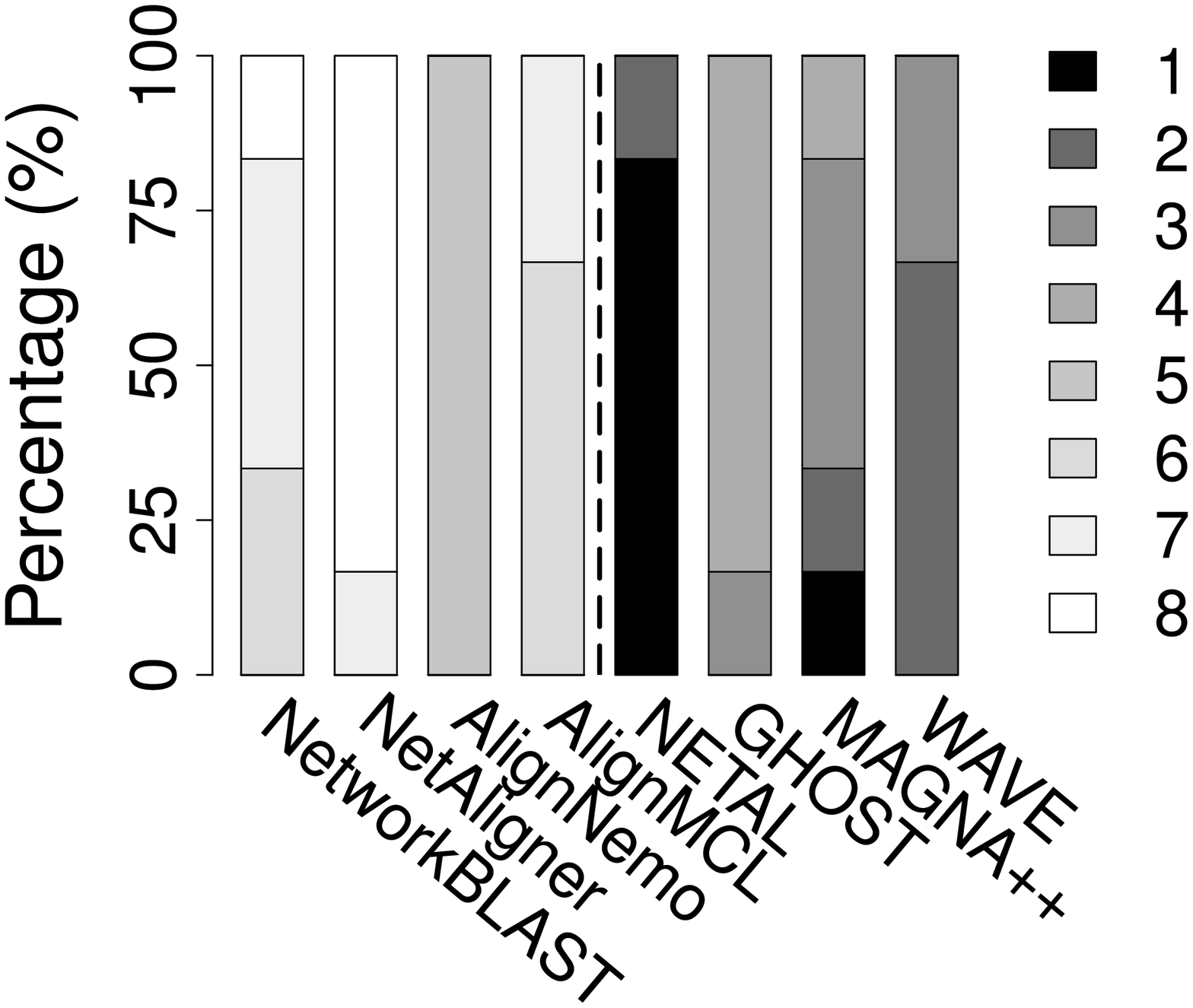}
(e)\includegraphics[width=0.21\linewidth]{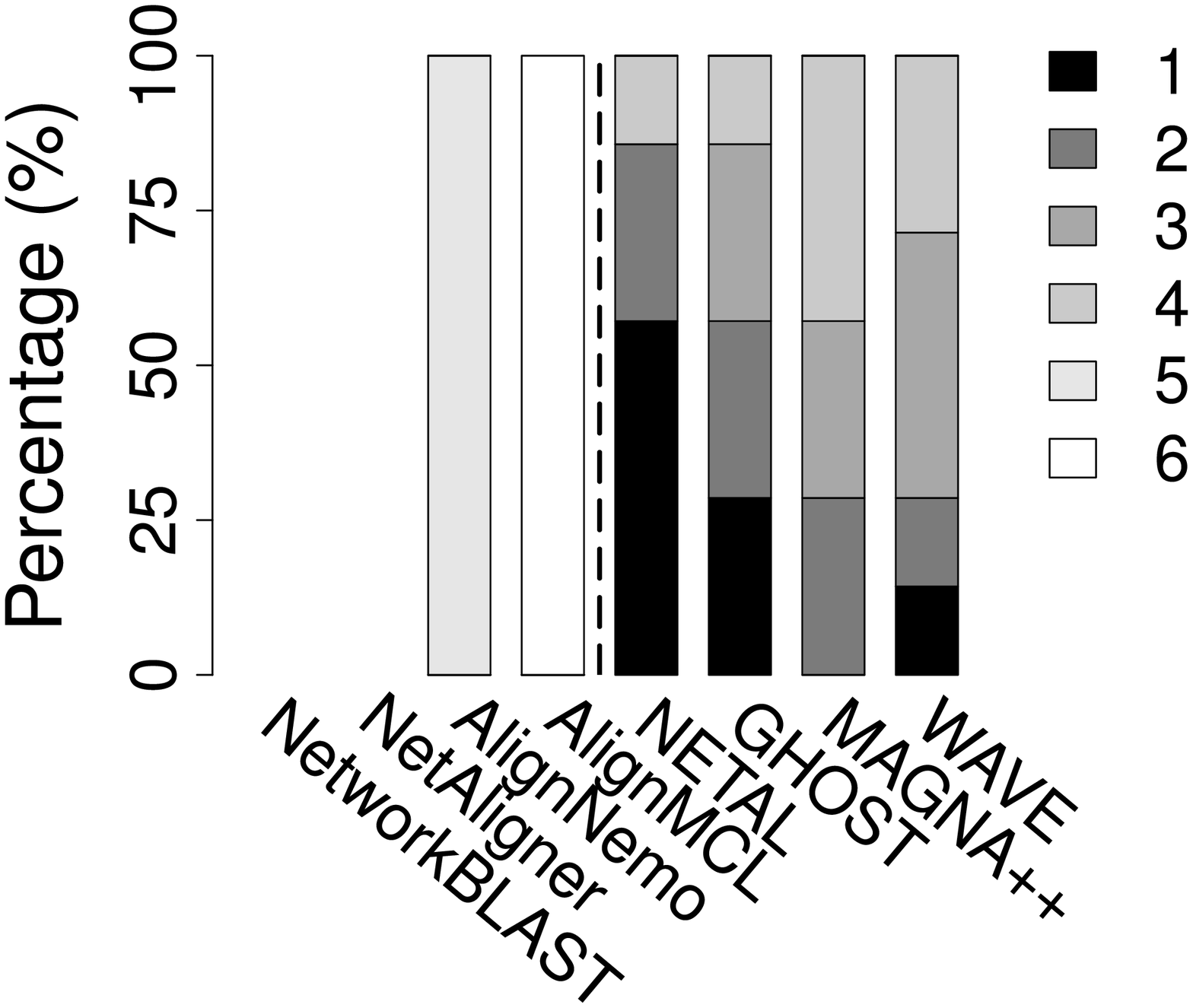}
(f)\includegraphics[width=0.21\linewidth]{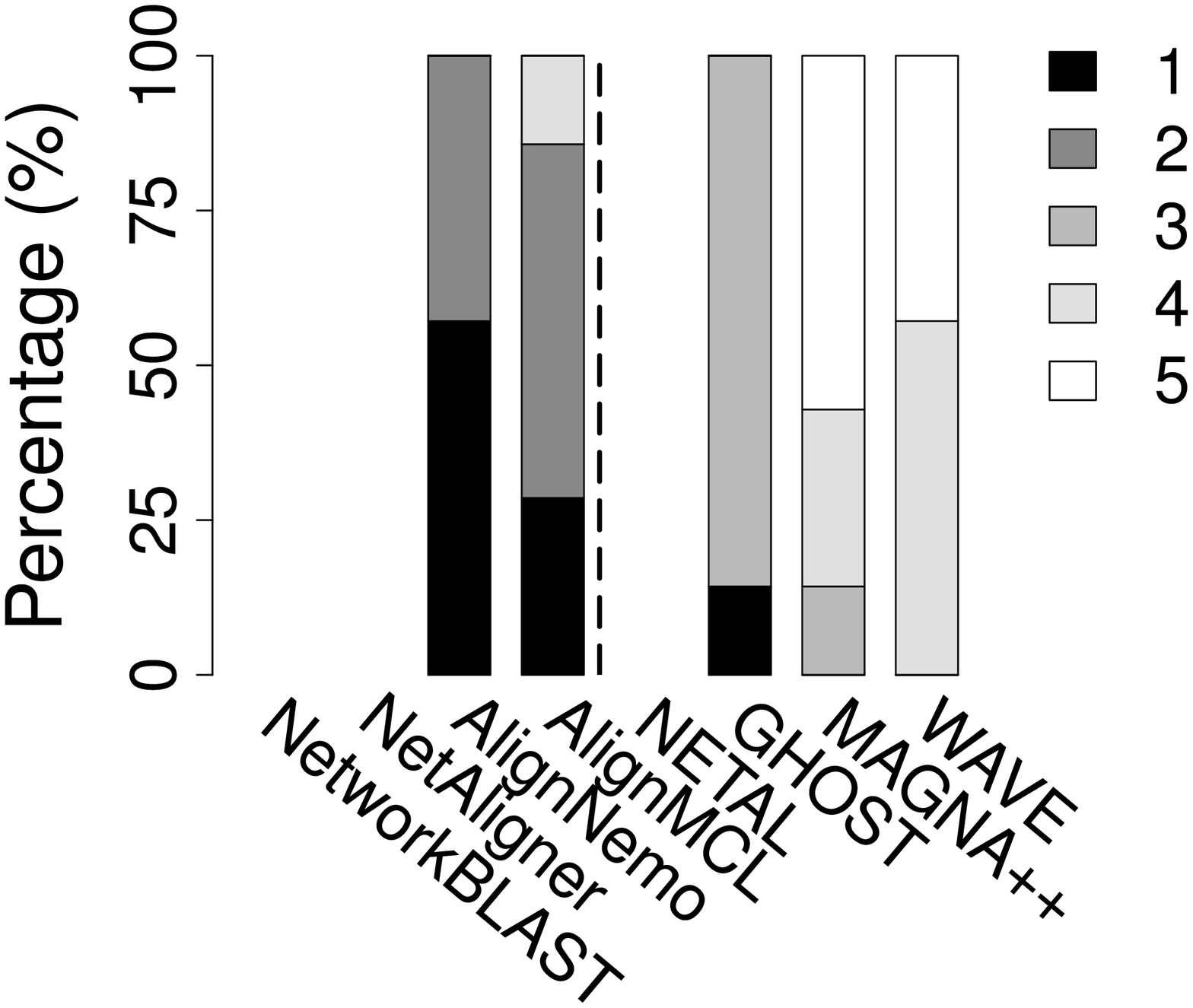}
(g)\includegraphics[width=0.21\linewidth]{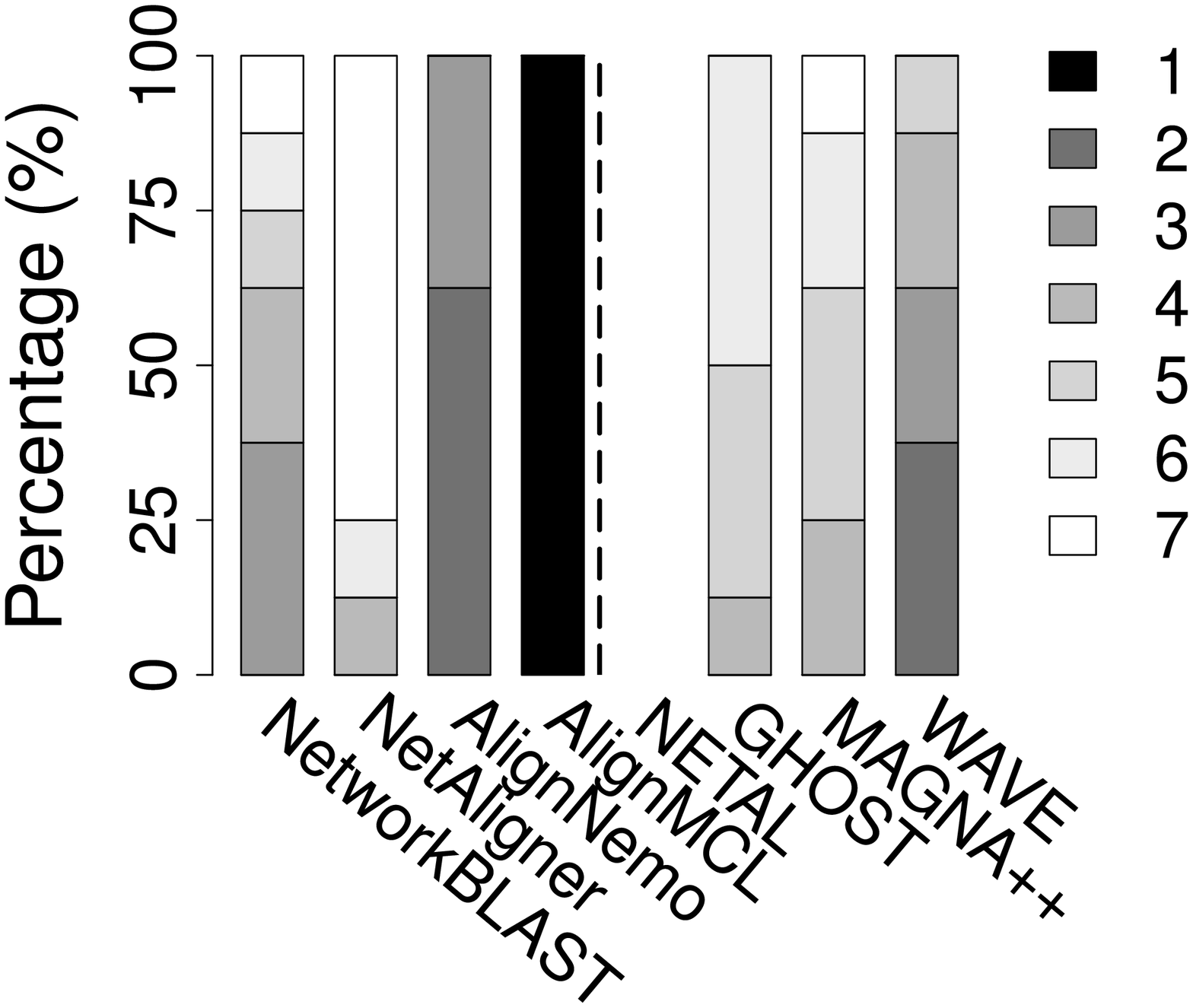}
(h)\includegraphics[width=0.21\linewidth]{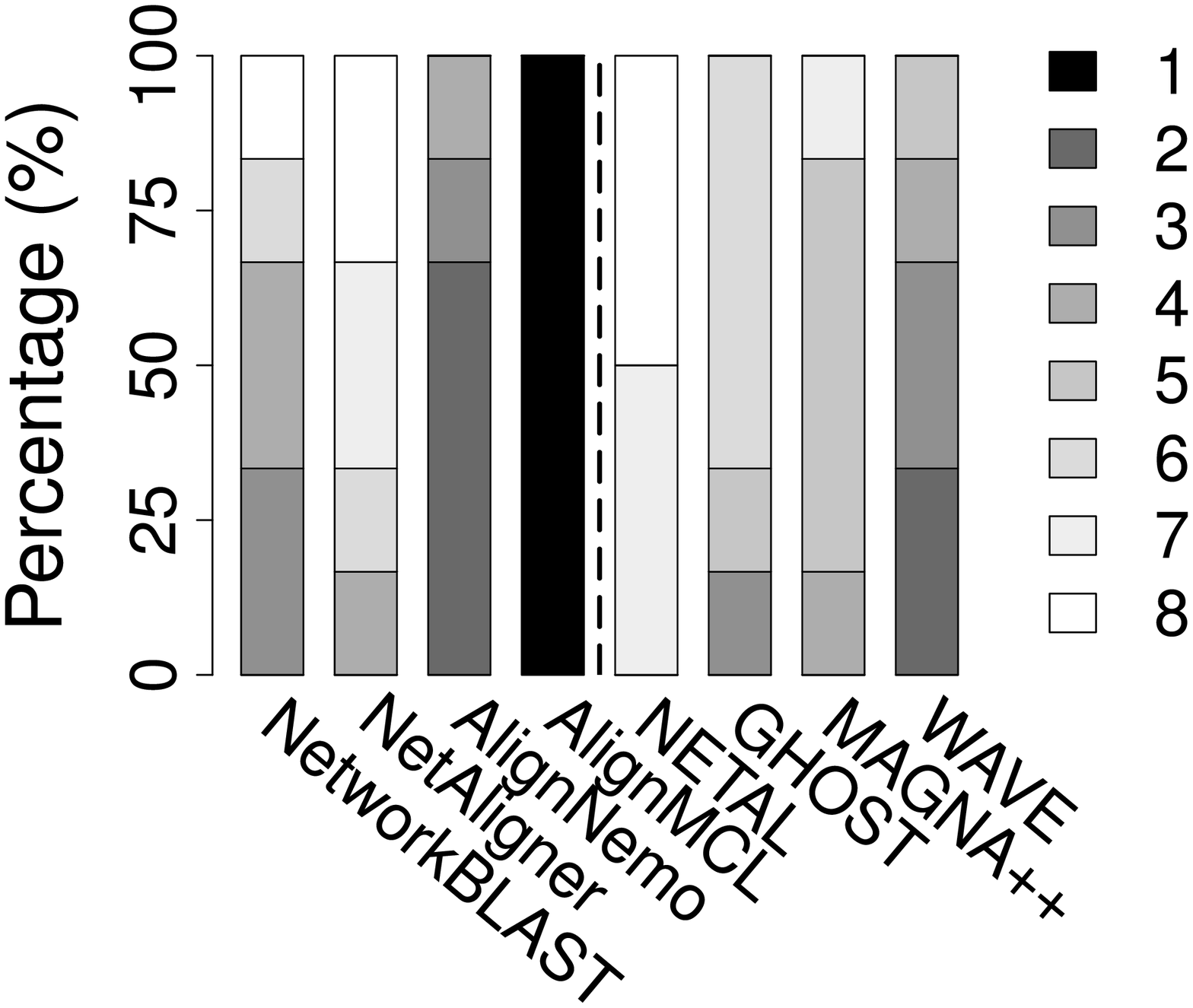}
\vspace{-0.3cm}
\caption{
Detailed comparison of LNA and GNA for networks with unknown true node
mapping from four different species (i.e., yeast, fly, worm and human)
containing four different types of PPIs (i.e., Y2H$_1$, Y2H$_2$,
PHY$_1$, and PHY$_2$) with respect to the topological NCV-GS$^3$
measure (panels
\textbf{(a)}-\textbf{(d)}) and the biological F-PF measure (panels \textbf{(e)} - \textbf{(h))},
for T (panels \textbf{(a)} and \textbf{(e)}), T\&S (panels
\textbf{(b)} and \textbf{(f)}), S (panels \textbf{(c)} and \textbf{(g)}), 
and B (panels \textbf{(d)} and \textbf{(h)}). Each bar shows the
percentage of the aligned network pairs for which the given NA method
performs as the $1^{st}$ best (1), $2^{nd}$ best (2), $3^{rd}$ best
(3), etc. among all of the NA (LNA or GNA) methods with respect to the
corresponding type of information used in NCF and the corresponding
alignment quality measure. A missing bar indicates that the given NA
method cannot use the corresponding type of information in NCF and
thus no result is produced. }
\label{some example}
\label{graph_biogrid_ranking_aligners}
\end{figure*}

Next, we zoom into the above results (Figure
\ref{graph_biogrid_ranking_aligners} and Supplementary Figures S8-S12) in order to identify the best NA method(s). 
Recall that for LNA, NetworkBlast and NetAligner do not allow for
using topological information in NCF. So, we cannot consider these
methods for T and T\&S. Given this, the results for LNA are as
follows. With respect to topological alignment quality: for T and
T\&S, and B, AlignNemo is the best method; for S, AlignMCL is the best
method. With respect to biological alignment quality: for T, AlignNemo
is the best method; for T\&S, AlignNemo and AlignMCL are the best
methods and are comparable, with slight superiority of AlignNemo; for
S and B, AlignMCL is the best method. Hence, we conclude AlignNemo and
AlignMCL to be the best of all analyzed LNA methods, of which under
the best-case scenario (B) AlignMCL is superior. Recall that for GNA,
NETAL does not allow for using sequence information in NCF. So, we
cannot consider this method for T\&S and S. (Clearly, for this method,
B is the same as T). Given this, the results for GNA are as
follows. With respect to topological alignment quality: for T, NETAL
is the best method; for T\&S, WAVE is superior; for S, MAGNA++ is the
best method; for B, NETAL is superior.  With respect to biological
alignment quality: for T, NETAL and GHOST are the best methods, with
slight superiority of NETAL over GHOST; for T\&S, GHOST is superior;
for S and B, WAVE is the best method. Hence, we conclude that for GNA,
in this analysis, the best method varies depending on whether we are
measuring topological versus biological alignment quality and
depending on the type of information used in NCF.

\subsubsection{Robustness to the choice of  network data}\label{sect:robust}
We aim to study the effect on results of using different network sets
(PHY$_1$, PHY$_2$, Y2H$_1$, and Y2H$_2$), in order to test the
robustness of the results to the choice of PPI type and confidence
level. We find that for each of the ``all methods'' and ``best method''
comparisons, topological and biological alignment quality, and T, T\&S,
S and B, results are consistent across the different network sets
(Supplementary Figure S13), and they are consistent with the above
reported results for all four network sets combined (Figure
\ref{graph_biogrid_overall_comparison}).  Thus, our NA evaluation
framework is robust to the choice of network data.

\subsubsection{Summary}

Overall, when using only topological information in NCF, GNA
outperforms LNA in terms of both topological and biological alignment
quality. When adding sequence information to NCF, GNA is superior in
terms of topological alignment quality, while LNA is superior in terms
of biological quality. The best of all LNA methods are AlignMCL and
AlignNemo.  The best of all GNA methods varies depending on whether
one is measuring topological versus biological alignment quality and
on the type of information used in NCF. Importantly, our evaluation
framework is robust to the choice of network data to be aligned.

The reason why GNA outperforms LNA in terms of \emph{topological}
alignment quality (meaning that GNA identifies larger amount of
conserved edges and larger conserved subgraphs that LNA), irrespective
of the type of NCF information used during the alignment construction
process, could be due to the following key difference between the
design goals of LNA and GNA. Namely, LNA aims to find \emph{small} (on
the order of a dozen nodes) but highly-conserved subnetworks,
irrespective of the overall similarity between the compared
networks. On the other hand, GNA aims to find a \emph{large} conserved
subgraph (though at the expense of matching local regions
suboptimally), and typically it does so by directly optimizing edge
conservation (and possibly other measures) while producing
alignments. As such, simply by design, GNA might have an advantage
over LNA in terms of the expected topological alignment quality, which
our results confirm.

In terms of \emph{biological} alignment quality, GNA again outperforms
LNA for T. This indicates that when using within NCF only biological
information encoded into network topology (i.e., when not using any
biological information external to network topology, such as sequence
information), GNA leads to better biological predictions than
LNA. Also, in this case, the topological alignment quality results
correlate well with the biological alignment quality results (as GNA
is superior to LNA in both cases). However, when some amount of
sequence information is included into NCF (corresponding to T\&S and
S), the topological alignment quality results do not correlate with
the biological alignment quality results (as GNA is superior in the
first case, while LNA is superior in the second case). The reason
behind LNA's superiority over GNA in terms of biological alignment
quality for T\&S and S could again be due to differences in their key
design goals. Namely, unlike GNA, LNA uses the notion of the alignment
graph to search for highly conserved subnetworks (Supplementary
Section S2). When sequence information \emph{is} used within NCF,
nodes in this graph contain sequence-based orthologs, i.e., highly
sequence-similar proteins from different networks. Since high sequence
similarity often corresponds to high functional similarity, and since
our measures of biological alignment quality are based on the notion
of functional similarity between aligned proteins, by design LNA is
``biased'' towards resulting in high biological quality whenever
sequence information is used in NCF. However, LNA fails to produce
biologically as good alignments as GNA when only topological
information is used in NCF, as discussed above.

\subsection{Running time method comparison}

The results from Sections \ref{sect:results_known} and
\ref{sect:results_unknown} compare the different methods in terms of
alignment accuracy. It is also important to compare the methods in
terms of computational complexity, which is the goal of this section.

We run all NA methods on the same Linux machine with 64 CPU cores (AMD
Opteron(tm) Processor 6378) and 512 GB of RAM.  Since some NA methods
(all LNA methods, as well as NETAL and WAVE GNA methods) can only run
on one core while the others (GHOST and MAGNA++ GNA methods) can run
on multiple cores, for fair comparison, we run all methods on a single
CPU core. An exception is GHOST, as its implementation still uses two
threads even when its code is configured to use one core.  We analyze
the methods' entire running times, both for computing node
similarities and for constructing alignments.  Also, we measure only
running times needed to construct alignments, ignoring the time needed
to precompute node similarities.  We do the above when aligning worm
and yeast PPI networks of Y2H$_1$ type (Table
\ref{table:time}). We choose these networks because both are
relatively small, and thus, the execution time for the slowest of all
methods on a single core is reasonable (within one day). For any other
network pair, running the slowest method on a single core would take
much longer.

Our findings are as follows. For the entire running time, overall, for
T, GNA is faster than LNA; for T\&S, GNA methods run similarly to LNA
methods. For S, LNA is faster than GNA.
For only the time needed to construct alignments, overall, LNA methods
run faster than GNA methods for each of T, T\&S, and S (Table
\ref{table:time} and Supplementary Section S6.3).  

In addition to the above single-core analysis, we give each method the
best-case advantage, by running the parallelizable methods (GHOST and
MAGNA++ GNA methods) on multiple cores; we use as many cores as
possible with the given method implementation, where 64 cores is the
maximum imposed by our machine. We show these results also in Table
\ref{table:time}, in parentheses.  As expected, running the two NA
methods on multiple cores indeed speeds up the methods' running
times. We do not necessarily see a linear decrease in running time
with the increase in the number of cores, as not all parts of the
given method are parallelizable.

Our results for the best-case, multi-core analysis are as follows. For
the entire running time, for T, GNA remains faster than LNA. However,
for T\&S and S, unlike in the above single-core analysis where LNA is
comparable or superior to GNA, GNA is now always comparable (if not
even superior) to LNA.
For only the time needed to construct alignments, LNA mostly remains
faster than GNA (Table \ref{table:time} and Supplementary Section S6.3).

\begin{table}[ht!]
\centering
\scriptsize
\begin{tabular}{|c|c|c|c|c|c|c|c|}
  \hline
  \multirow{3}{*}{Type} &   \multirow{3}{*}{Method} &  \multicolumn{3}{c|}{\multirow{2}{*}{Entire time (min)}}  & \multicolumn{3}{c|}{Only time needed to }\\ 
     &    &  \multicolumn{3}{c|}{}  & \multicolumn{3}{c|}{construct alignments (min)}\\ \cline{3-8}
    &  & T  &  T\&S & S & T  &  T\&S & S\\\hline
   \multirow{4}{*}{LNA} & NetworkBLAST &  - & - & 372.6 &  - & - & 7.3\\  \cline{2-8}
   & NetAligner & - & - & 368.2& - & - & 2.35\\ \cline{2-8}
   & AlignNemo &  375.5 & 450.3 & 370.0&  4.9 & 0.4 & 0.4\\  \cline{2-8  }
   & AlignMCL & 377.0 & 452.1 & 365.2& 1.6 & 1.75 & 1.7\\ \hline
   \multirow{4}{*}{GNA}  & NETAL &  0.4 & - & -&  0.4 & - & -\\  \cline{2-8}
   & GHOST & 78.2 & 438.5 & 435.3&7.5 & 9.5 & 10.7\\  
   &  & (16.8) & (381.8) & (378.1)& (4.2) & (6.5) & (6.4)\\  \cline{2-8}
   & MAGNA++ & 287.8 & 768.9 & 690.7&  224.6 & 225.2 & 221.7\\   
   & & (31.6) & (474.4) & (383.4)& (14.7) & (14.3) & (14.1)\\  \cline{2-8}
   & WAVE & 17.15 & 450.8 & 369.7& 2.9 & 3.1 & 2.8\\ \hline
\end{tabular}
\caption{Representative running time comparison of the different NA methods, 
for each of T, T\&S, and S. Both the entire running times and only the
running times for computing alignments are shown. Values outside
parentheses are for the single-core analysis, while values in
parentheses are for the multi-core analysis (when applicable). The `-'
character indicates that the given method cannot use the corresponding
type of information in NCF and thus no result is produced.}
\label{table:time}
\end{table}

\subsection{Novel protein function predictions}

Finally, we contrast LNA against GNA in the context of learning novel
protein functional knowledge.  We identify alignments in which the
aligned network regions are significantly functionally similar
according to known functional knowledge. Then, from such alignments,
we predict novel functional knowledge in currently unannotated network
regions whenever such regions are aligned to functionally annotated
network regions (Section
\ref{sect:novel_predictions}).

\begin{figure}[h!]
\centering
(a)\includegraphics[width=0.2\linewidth]{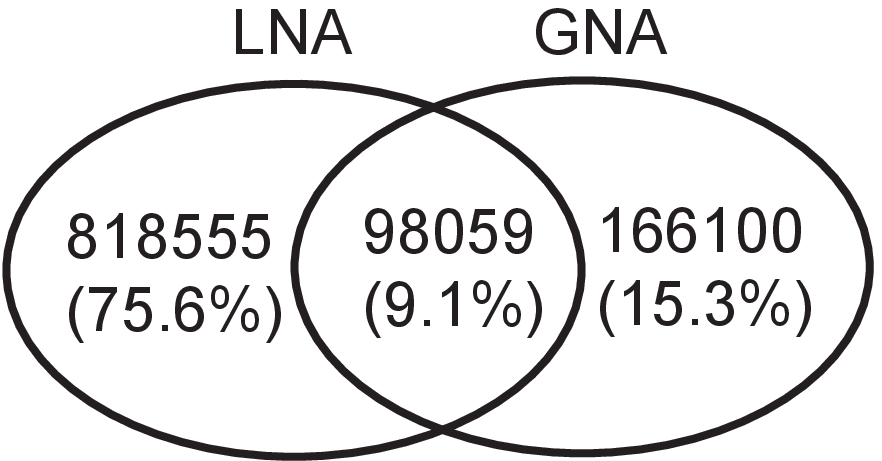}
(b)\includegraphics[width=0.2\linewidth]{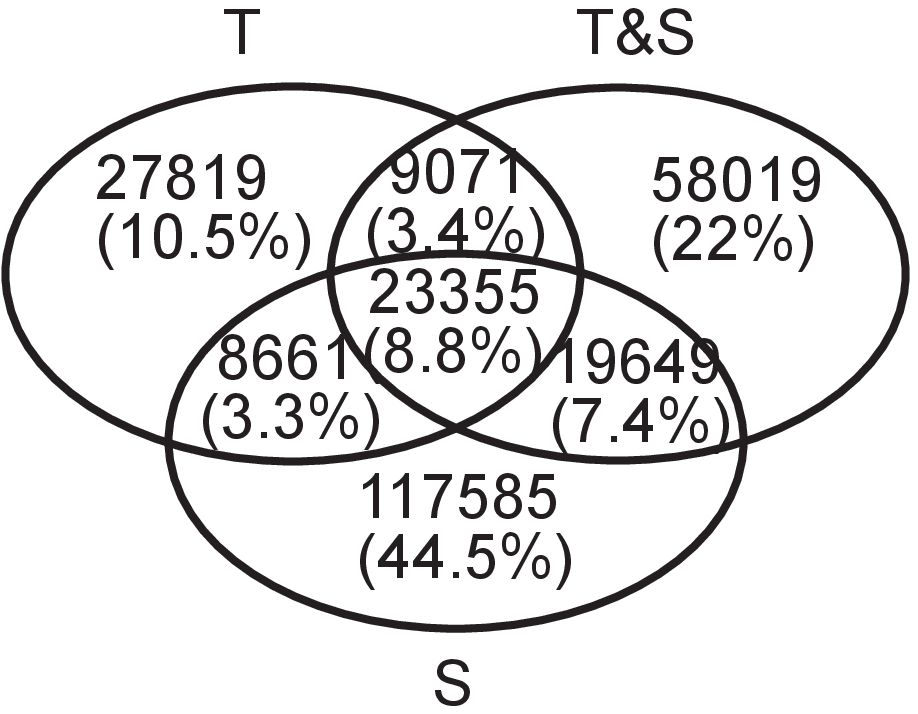}
\vspace{-0.3cm}
\caption{Overlap of unique novel protein function predictions between  
\textbf{(a)} LNA and GNA over all of T, T\&S, and S combined, \textbf{(b)} 
T, T\&S, and S for GNA.  See Supplementary Figure S14 for overlap of unique novel protein function predictions between T, T\&S, and S for LNA. }
\label{some example}
\label{graph:overlap}
\end{figure}

We find that LNA and GNA produce very different predictions,
indicating their complementarity when learning new knowledge.  Of the
predictions made by all (LNA or GNA) methods for all of T, T\&S, and
S, significant portion come from LNA only or GNA only, and only 9.1\%
come from both LNA and GNA (Figure
\ref{graph:overlap} (a)).

We zoom into the above results for each of LNA (Supplementary Figure
S14) and GNA (Figure
\ref{graph:overlap} (b)) to study the effect on the prediction
results of using different types of information in NCF. We aim to test
whether using some amount of topological information in NCF
(corresponding to T or T\&S) can yield unique predictions that are not
captured when using only sequence information in NCF (corresponding to
S). If so, this would confirm that additional biological knowledge is
encoded in network topology compared to sequence data. Indeed, this is
what we observe, for each of LNA and GNA: most predictions are unique
to the different types of NCF information. Thus, network topology and
sequence information complement each other when learning new
biological knowledge.

\section{Conclusions}

In this paper, we systematically evaluate LNA against GNA. Our
findings provide guidelines for researchers to properly demonstrate
the superiority of a newly proposed NA (LNA or GNA) method. That is,
we recommend that researchers evaluate the topological quality of a
new NA method against state-of-the-art GNA (rather than only LNA)
methods, irrespective of the type of information used in NCF, and that
they evaluate the biological alignment quality of the new NA method
against state-of-the-art GNA (rather than only LNA) methods when only
T is used in NCF and against LNA (rather than only GNA) methods when S
is also used in NCF.  NA can be used to complement the across-species
transfer of functional knowledge that has traditionally relied on
sequence alignment.

\section*{Acknowledgement}

This work was funded by the National Science Foundation [CAREER
CCF-1452795, CCF-1319469, and IIS-0968529].

\bibliographystyle{plain}
\bibliography{paper}

\end{document}